\documentclass[usegraphicx,usedcolumn,usenatbib,epsf,graphics]{mn2e}
\usepackage{rotating}
\usepackage{amssymb}
\newif\ifAMStwofonts
\AMStwofontstrue

\ifCUPmtlplainloaded \else
  \ifAMStwofonts \else 

  \fi
\fi

\title[Pre-main sequence stars in the Lagoon Nebula (M8)]{Pre-main sequence 
stars in the Lagoon Nebula (M8)\thanks{This paper includes data gathered with the 6.5 meter Magellan Telescopes located at Las Campanas Observatory, Chile.}}

\author[Arias et al.]
{
J. I. Arias$^1$\thanks{Fellow of CONICET, Argentina},
R. H. Barb\'a$^2$\thanks{Member of Carrera del Investigador Cient\'{\i}fico, CONICET, Argentina},
N. I. Morrell$^{3}$
\\
$^1$ Facultad de Ciencias Astron\'omicas y Geof\'{\i}sicas, Universidad Nacional de La Plata, Paseo del Bosque S/N, B1900FWA La Plata, Argentina\\
$^2$ Departamento de F\'{\i}sica, Universidad de La Serena, Benavente 980, 
La Serena, Chile\\
$^3$ Las Campanas Observatory, Carnegie Observatories, Casilla 601, 
La Serena, Chile\\
}

\pubyear{2003}

\begin{document} 

\maketitle

\begin{abstract}
We report the discovery of new pre-main sequence (PMS) stars in 
the Lagoon Nebula (M8) at a distance of 1.25~kpc, 
based on intermediate resolution spectra obtained with the Boller \& Chivens 
spectrograph at the 6.5\,m Magellan I telescope (Las Campanas Observatory, 
Chile). 
According to the spectral types, the presence of emission lines and the 
lithium $\lambda$\,6708 absorption line, we are able to identify 27 classical 
T~Tauri stars, 7 weak-lined T~Tauri stars and 3 PMS emission objects with 
spectral type G, which we include in a separated stellar class denominated
``PMS Fe/Ge class''. 
Using near-infrared photometry either from 2MASS or from our own previous work
we derive effective temperatures and luminosities for these stars 
and locate them in the Hertzsprung-Russell diagram,
in order to estimate their masses and ages. We find that almost all of our 
sample stars are younger than $3\times10^6$ years and span over a range of 
masses between 0.8 and 2.5~$M_{\odot}$.
A cross-correlation between our spectroscopic data and the X-ray
sources detected with the {\em Chandra}~ACIS instrument is also presented.
\end{abstract}


\section{Introduction}
The Lagoon Nebula (Messier 8 = NGC~6523 - NGC~6530) is an 
extended H\,{\sc ii} region in the Galaxy. It is immersed in a giant
molecular cloud which extends to the very young open cluster NGC~6530.
The UV light responsible for the ionization of the M8 nebula primarily 
originates in three O-type stars of NGC~6530: 
the massive binary systems 9\,Sagitarii, O4\,V\,{\small ((f))}, 
and HD~165052, O6.5\,V\,{\small ((f))} + O7.5\,V\,{\small ((f))}, 
and the extremely young object Herschel~36, O7.5\,V\,{\small ((f))}.
The strong UV radiation from the latter star has developed a
distinctive bipolar blister-type H\,{\sc ii} region known as the Hourglass 
Nebula.
Additionally, NGC~6530 contains more than 60 B-type
stars, which makes it ~3-4 times richer in massive stars than the Orion 
Nebula cluster. In contrast to what is observed in other star forming regions, 
the O stars in NGC~6530 are curiously located in the cluster periphery rather 
than toward its centre. 
The low-mass stars coeval to these massive OB stars are 
expected to still be in their pre-main sequence evolutionary stages.

The previoulsy known stellar population of NGC~6530 shows evidence of 
either recent or on-going star forming processes.
Lightfoot et al. (1984) proposed the existence of three different stellar 
generations within this cluster; consequently it may provide an example of 
the sequential star formation mechanism (Lada et al. 1976).
Several optical studies have been devoted to this region since the first one
by Walker (1957). 
Van den Ancker et al. (1997) studied 
the probable members of NGC~6530 and concluded that the process of star 
formation in this cluster must have started a few times $10^7$ years ago and 
that, for the less massive stars, is probably still going on today.
Based on $UBVRI$ and H$\alpha$ photometry, Sung et al. (2000) found several 
pre-main sequence (PMS) candidates and
estimated for the cluster an age of 1.5 million years.
After that, Prisinzano et al. (2005) performed $BVI$ photometry down to
$V=22$ on NGC~6530. They correlated their optical catalog with the list of
X-ray sources detected with the {\em Chandra} satellite and published by 
Damiani et al. (2004), 
finding more than 800 objects in common, 90\% of which could be PMS stars.

More recently, Arias et al. (2006) investigated the Hourglass Nebula in
near-infrared (NIR) wavelenghts and detected almost 100 NIR excess sources 
identifiable of young stellar objects (YSOs), such as Class~I ``protostars'',
T~Tauri and Herbig~Ae/Be stars. 
Moreover, using archival {\em HST} images, they found four Herbig-Haro 
objects for the first time in the region.

\begin{table*}
\begin{minipage}{155mm}
\caption{Data extracted from literature for the 46 observed objects in M8. See text (Sec.~\ref{select}) for more detail.}
\label{literature}
\begin{tabular}{clcccccccc}
\hline\\
ABM & SCB/2MASS & $\alpha$(J2000) & $\delta$(J2000) & $V$ & $K_s$ & $J-H$ & $H-K_s$ & WFI & DFM2004 \\
  $[1]$   &  ~~~~~~~[2]   &        [3]  &        [4]    &  [5]  &  [6]    &   [7]   &   [8]     &  [9]  &  [10] \\
\hline\\
 1 & 2MASS-270.894257  & 18:03:34.6 & -24:22:17.7 &   ...   & 14.931     & 0.380 & -0.191 & ...   & ... \\  
 2 & SCB 106           & 18:03:35.3 & -24:22:26.5 &  14.432 & 11.126     & 0.493 &  0.978 & 19640 & ... \\ 
 3 & SCB 1031          & 18:03:37.9 & -24:21:42.0 &  17.985 & 11.603$^*$ & 1.105 &  0.838 & ...   & ... \\ 
 4 & SCB 1040          & 18:03:38.6 & -24:22:23.9 &  18.476 & 11.639$^*$ & 1.316 &  1.019 & ...   & ... \\ 
 5 & SCB 1036          & 18:03:38.8 & -24:22:34.8 &  18.891 & 11.614$^*$ & 1.323 &  0.561 & ...   & ... \\ 
 6 & SCB 1018          & 18:03:39.5 & -24:23:00.9 &  18.758 & 11.610$^*$ & 1.450 &  0.856 & ...   & ... \\ 
 7 & SCB 1035          & 18:03:39.9 & -24:23:02.5 &  19.156 & 12.176$^*$ & 1.295 &  0.623 & ...   & ... \\ 
 8 & SCB 1032          & 18:03:40.3 & -24:22:03.5 &  18.722 & 10.955$^*$ & 1.344 &  0.614 & ...   & ... \\ 
 9 & 2MASS-270.918889  & 18;03:40.5 & -24:23:31.9 &   ...   & 11.152$^*$ & 1.445 &  0.840 & ...   & ... \\  
10 & SCB 146           & 18:03:40.7 & -24:23:16.3 &  15.638 & 10.174$^*$ & 1.253 &  0.643 & 18270 & ... \\ 
11 & SCB 148 - KS\,2   & 18:03:41.1 & -24:22 41.3 &  15.039 & 11.040$^*$ & 0.882 &  0.285 & ...   & ... \\ 
12 & SCB 174           & 18:03:43.7 & -24:23 39.8 &  16.898 & 11.942$^*$ & 1.043 &  0.466 & 17527 & ... \\ 
13 & SCB 184           & 18:03:45.2 & -24:23 25.2 &  16.531 & 10.430$^*$ & 1.305 &  0.925 & 17998 & ... \\ 
14 & SCB 201           & 18:03:47.0 & -24:23 09.0 &  12.601 & 10.718     & 0.331 &  0.072 & 18472 & ... \\ 
15 & SCB 240 - LkH$\alpha$~108 & 18:03:50.8 & -24:21 10.9 &  11.722 &  8.363   & 0.858 &  0.851 & 21869 &  15 \\ 
16 & SCB 284           & 18:03:57.1 & -24:17 00.4 &  15.821 & 12.230     & 0.661 &  0.354 & 28677 &  ND \\ 
17 & SCB 292           & 18:03:58.3 & -24:16 49.1 &  16.837 & 11.530     & 1.016 &  0.712 & 29045 &  54 \\ 
18 & SCB 388           & 18:04:07.9 & -24:23 11.6 &  16.699 & 12.266     & 0.793 &  0.383 & 18388 & 151 \\ 
19 & SCB 390           & 18:04:07.9 & -24:23 12.7 &  15.583 & 11.784     & 0.793 &  0.383 & 18365 & 151 \\ 
20 & SCB 410           & 18:04:09.7 & -24:27 10.3 &  16.688 & 11.464     & 0.783 &  0.564 & 12983 & 177 \\ 
21 & SCB 418           & 18:04:10.3 & -24:23 22.8 &  16.505 & 11.584     & 0.759 &  0.344 & 18064 & 191 \\ 
22 & SCB 422           & 18:04:10.6 & -24:26 56.0 &  15.532 & 10.481     & 0.990 &  0.616 & 17814 & 194 \\ 
23 & SCB 425           & 18:04:11.1 & -24:23 30.8 &  15.087 & 7.664      & 1.297 &  0.495 & 17814 &  ND \\ 
24 & SCB 440           & 18:04:12.5 & -24:19 43.0 &  15.985 & 11.952     & 0.744 &  0.134 & 24550 & 216 \\ 
25 & 2MASS-271.061044  & 18:04:14.6 & -24:19:03.4 &  18.410 & 12.956     & 0.876 &  0.319 & 25568 & 257 \\
26 & SCB 482           & 18:04:15.8 & -24:19 01.6 &  15.448 & 11.299     & 0.800 &  0.407 & 25592 & 285 \\ 
27 & SCB 486           & 18:04:16.0 & -24:18 46.2 &  16.904 & 11.638     & 1.106 &  0.806 & 25944 & 288 \\ 
28 & SCB 493           & 18:04:16.4 & -24:25 03.2 &  16.234 & 12.849     & 0.496 &  0.825 & 15291 & ND \\ 
29 & SCB 495           & 18:04:16.4 & -24:24 39.1 &  16.730 & 11.100     & 1.096 &  0.781 & 15923 & 297 \\ 
30 & SCB 508 - LkH$\alpha$~111 & 18:04:17.5 & -24:19 09.4 &  16.732 & 10.385   & 1.567 &  0.982 & 25513 & 311 \\ 
31 & SCB 531           & 18:04:19.4 & -24:22 54.7 &  16.029 & 11.027     & 0.939 &  0.470 & 18856 & 345 \\ 
32 & SCB 540           & 18:04:20.1 & -24:22 48.2 &  16.172 & 11.257     & 0.886 &  0.833 & 19052 & 363 \\ 
33 & SCB 547           & 18:04:20.4 & -24:28 19.6 &  14.935 & 10.909     & 0.779 &  0.230 & 11699 & 368 \\ 
34 & SCB 552           & 18:04:20.8 & -24:28 02.7 &  15.389 & 10.186     & 0.894 &  0.652 & 12044 & 387 \\ 
35 & SCB 666           & 18:04:29.3 & -24:23 43.3 &  16.528 & 11.735     & 0.914 &  0.500 & 17441 & 584 \\ 
36 & SCB 681           & 18:04:30.9 & -24:26 34.9 &  15.000 & 11.338     & 0.720 &  0.262 & 13569 & 618 \\
37 & 2MASS-271.128826 A & 18:04:34.9 & -24:26:43.2 & 18.916 & 11.332     & 1.152 &  0.825 & 13430 & 617$^\dag$ \\ 
38 & 2MASS-271.128826 B & 18:04:34.9 & -24:26:43.2 &   ...   &  ...      & ...   & ...    & ...   & ... \\ 
39 & 2MASS-271.145562   & 18:04:34.9 & -24:26:24.0 &   ...   & 13.796    & 0.663 & 0.277  & ...   & ND  \\
40 & SCB 726            & 18:04:35.1 & -24:26 12.9 & 16.884  & 11.974    & 1.152 & 0.783  & 13943 & ND \\ 
41 & 2MASS-271.164608   & 18:04:39.5 & -24:17:20.1 & 17.645  & 11.376    & 1.303 & 0.528  & 28075 & 732 \\
42 & SCB 782            & 18:04:40.9 & -24:17 11.2 & 16.253  & 11.000    & 1.053 & 0.698  & 28348 & 748 \\ 
43 & SCB 810            & 18:04:43.6 & -24:27 38.8 & 16.253  & 10.023    & 1.214 & 0.759  & 12467 & 769 \\ 
44 & SCB 862            & 18:04:48.6 & -24:26 40.7 & 16.897  & 10.064    &
1.578 & 0.954  & 13474 & ND \\ 
45 & SCB 879 - LkH$\alpha$~115  & 18:04:50.6 & -24:25 42.5 & 11.942  &  9.728  & 0.291 & 0.434  & 14474 & 828 \\ 
46 & 2MASS-271.215590   & 18:04:51.7 & -24:25:47.5 &  ...    & 13.445    & 0.543 & 0.148  & ...   & ... \\ 
\hline\\
\end{tabular}
\footnotesize $^\ast$ Values from Arias et al. (2006).\\
$^\dag$ Being 2MASS-271.128826 a binary star with 
$1\farcs65$-separated components, the data from literature have been assigned 
to the brightest object.\\
\end{minipage}
\end{table*}

In spite of the extensive body of literature on M8 and NGC~6530, 
so far most of the stellar population members of this region lack of any 
spectroscopic observation.
This paper presents the results of a spectroscopic study 
of PMS candidates confirming  that  NGC\,6530 is 
an active star forming region with a rich population of YSOs.

\section{Observations}

\subsection{Target selection}
\label{select}

The selection of candidates for the spectroscopic follow-up was mainly based
on the lists of PMS candidates with H$\alpha$ emission in the optical 
photometric work by Sung et al. (2000). 
A minor group of targets in the Hourglass Nebula region were selected 
according to their position in the $JHK_s$ colour-colour diagram 
(Arias et al. 2006), where
they show near-infrared excess emission characteristic of circumstellar disks. 
Finally, a few objects with particular morphological appearance in {\em HST} 
images, such as extended or knotty shape and bow shock arcs around them 
(Arias et al. 2006), were also included in the sample.

In Table~\ref{literature} we list our sample stars along with their
coordinates and photometric data from the literature. Running source numbers
are in column 1 and identifications either from Sung et al. (2000) or from 
the 2MASS catalog (Skrutskie et al. 2006) are in column 2. 
Historical identifiers referring to emission
stars have been also added in a few cases. Columns 3 and 4 contain right
ascension and declination (J2000). $V$ magnitudes from Sung et al. (2000) or 
Prisinzano et al. (2005) are listed in column 5, while $K_s$ magnitudes 
and $J-H$, $H-K_s$ colours are listed in columns 6, 7 and 8, respectively. 
Infrared data were extracted from 2MASS for all the objects, except those
marked with an asterisk for which we used data from Arias et al. (2006). 
Column~9 and 10 contain cross-correlated identifications with
the optical sources in Prisinzano et al. (2005) (with nomenclature ``WFI'') 
and with the X-ray sources detected with the {\em Chandra} ACIS instrument 
and published by Damiani, Flaccomio, Micela et al. (2004) (with nomenclature 
``DFM2004''). ``ND'' refers to the non-detected sources in the FOV of the 
{\em Chandra} observation.

\subsection{Spectroscopy}

The observations were obtained at the 6.5\,m Magellan\,{\sc i} telescope 
(Baade) with the Boller \& Chivens spectrograph during the nights of July
29 and 30, 2003.
We used the Marconi 2048 $\times$ 515 CCD and the 600 l\,mm$^{-1}$ grating,
obtaining a dispersion of 1.6 \AA\,px$^{-1}$ over the wavelength 3850-7000~\AA.
The spatial scale on the detector is 0\farcs25 px$^{-1}$ and the slit length 
$72''$. Observing conditions were optimum, with the seeing ranging 
between 0\farcs6 and 1\farcs0.
The exposure times ranged from 30 to 1200 sec.
The slit was rotated conveniently in order to observe two or more objects
simultaneously.
The usual sets of bias and flat field calibrations were also obtained for each 
night along with several spectrophotometric standards to get flux calibrated 
spectra.
The data were processed and analysed with IRAF\footnote{IRAF is distributed 
by NOAO, operated by AURA, Inc., under agreement with NSF.} routines 
at La Serena University (Chile). Typical signal-to-noise ratios are 
$S/N\sim 50-200$.
The spectra were reduced in long slit mode in order to 
achieve a detailed analysis of the nebular material.
Special care was taken with the sky background subtraction
since most of the objects are deeply embedded in the nebula.
For some objects, in particular those located in the Hourglass region, 
the background emission was so intense and variable that subtracting it 
accurately all over the wavelenght range became extremely hard.
Because of this, some residua or artifacts might be present in any of the 
spectra.

\section{Results}

\subsection{Spectral classification}
\label{clasifica}

\begin{figure*}
\includegraphics[width=120mm, angle=90]{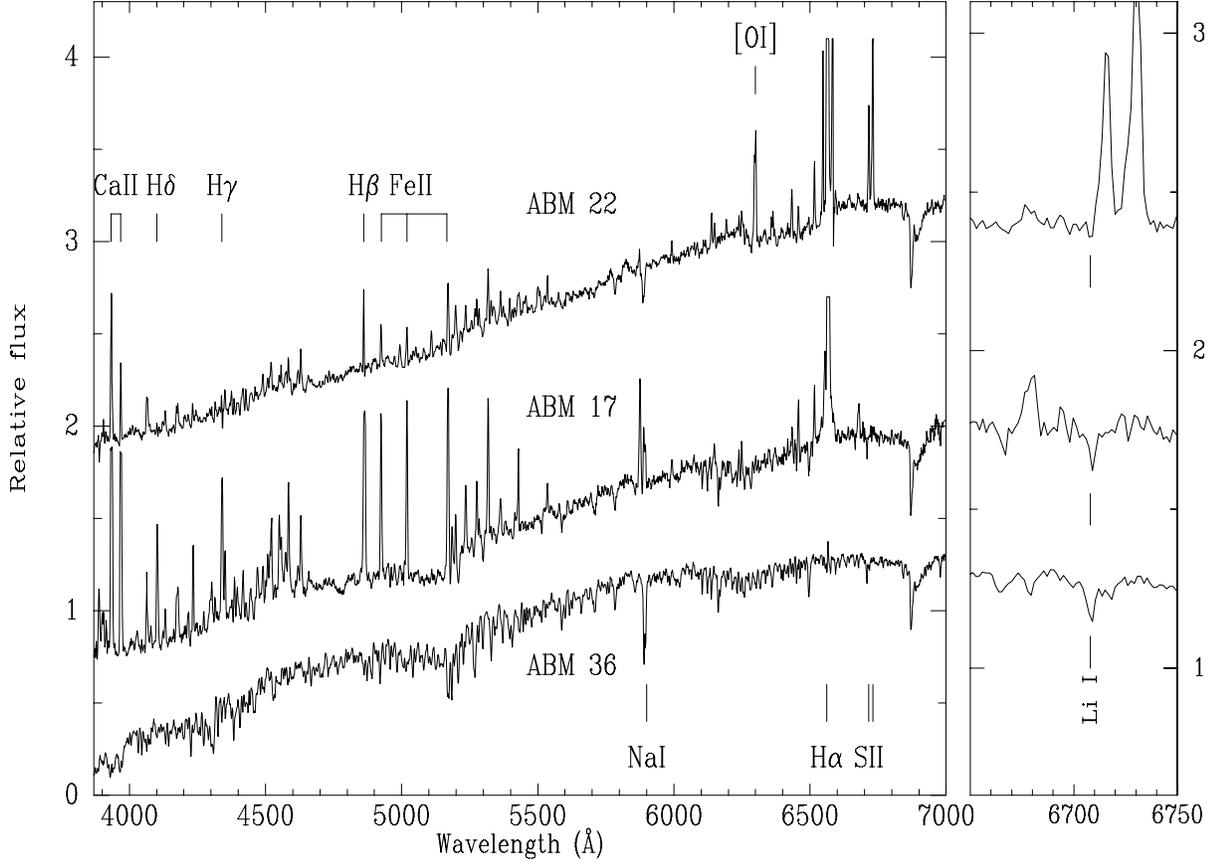}
\caption{Left-hand panel: sample spectra of the new PMS~stars observed in M8.
From top to bottom, a PMS~Fe/Ge object (ABM\,22), a CTT star (ABM\,17) and a 
WTT star (ABM\,36) are shown.
The spectra have been normalized to the continuum at $\lambda=5500$\,\AA\,and 
vertically shifted for clarity. In ABM\,22 and ABM\,17, the strongest emission 
lines have been truncated with the same purpose.
The rest-wavelenghts of some the most 
prominent lines, including the Balmer series (H$\delta$, H$\gamma$, H$\beta$, 
H$\alpha$), the Ca\,{\sc ii}~H\&K lines, the Fe\,{\sc ii} (42) multiplet, 
the Na\,{\sc i}\,$\lambda\lambda$5890, 5896 and the 
[S\,{\sc ii}]\,$\lambda\lambda$6716, 6731 lines are indicated. 
Right-hand panel: an enlargement of the same spectra showing the 
$6650-6750$\,\AA\,region. The ticks indicate the rest-wavelenght of the 
Li\,{\sc i}\,$\lambda$6708 line.}
\label{sample}
\end{figure*}

We have used the spectrophotometric standards of both Jacoby et al.'s (1984) 
and Pickles' (1998) spectrum libraries in order to determine the spectral 
types of the observed objects, although the resolution of the latter database 
is lower than that of our spectra.
The wide wavelength range of our spectra includes a reasonable number of 
features suitable for spectral typing. The classification was performed
by visual inspection of the spectra.
In the 3800-5600\,\AA\ region, we mainly considered the classification 
criteria described in the digital atlas by R.~O. 
Gray\footnote{http://nedwww.ipac.caltech.edu/level5/Gray/frames.html}.
Useful features for classifying late-type stars in the 5000-7000\,\AA\ 
region are the 
Mg\,{\sc i} lines at 5164-5173\,\AA, Na\,{\sc i} at 5890-5896\,\AA, 
Ca\,{\sc i} at 6162\,\AA\, and CaH at 6496\,\AA, 
as well as the TiO bands  at 5167, 5862 and 6159\,\AA.  
All the spectra were classified by the three authors independently. After
confronting the individual results, a final average spectral type was 
assigned to each object.

It is a well known fact that the photospheric absorption features in the
spectra of these young stars are veiled by the blue continuum excesses 
produced by the accretion mechanism. 
Moreover, in the most extreme objects, the photospheric continuum may appear 
completely hidden underneath 
a rich emission line spectrum originating in the circumstellar material.
Consequently, the accuracy of the spectral classification strongly depends on 
each case, ranging from less than $\pm$ 1 subclass 
to perhaps 2-3 subclasses for the spectra most severely affected 
by the accretion processes.
Most of our objects are late-type stars showing G or K spectral type along
with prominent H$\alpha$ emission. The presence of forbidden emission lines,
the Ca\,{\sc ii}~H\&K lines in emission and the Li\,{\sc i}~6708\,\AA\, 
absorption line, a  primary indicator of youth, 
are also observed in almost all targets, establishing the definitive 
PMS nature of these stars. 

Results of the spectroscopy of these PMS stars are shown in Table~\ref{pms}. 
Historically, the photospheric emission of T~Tauri stars with spectral types 
earlier than M0, has been successfully represented by the spectra of 
normal dwarf stars (luminosity class V) (e.g. Basri \& Batalha 1990). 
At cooler spectral types significant deviations between the spectra of 
T~Tauri and dwarf stars appear, as a consequence of the gravity-sensitive
molecular absorption bands that dominate the spectra (Torres-Dodgen \& 
Weaver 1993).
We find that only a minor fraction of the objects in our sample 
are best matched by  giant-like spectra (luminosity class III), 
whereas the majority are well reproduced by standard dwarf spectra, 
in good accordance with the mentioned results.
The luminosity class that best represents the photospheric emission of each 
star is also indicated in Table~\ref{pms}, next to the assigned spectral type.

In addition to the adopted 
spectral types and luminosity classes we present the equivalent widths of 
the H$\alpha$ and the Li\,{\sc i} lines and indicate additional emission 
lines observed in the spectra. In our convention, a negative equivalent width 
means an emission line. The errors quoted in parentheses refer to the  
repeatability of the measurements. Real errors of the Li\,{\sc i} equivalent 
widths may probably be larger due to the blending with neighbouring lines such 
as Fe\,{\sc i}~$\lambda$6705 and Ca\,{\sc i} $\lambda$6710.
It has been noted that for some objects which are deeply embedded in the 
intense highly variable nebula, slight variations ($\sim5~\%$) 
in the  background 
subtraction lead to large changes ($25-100~\%$) in the equivalent width 
of H$\alpha$ [W(H$\alpha$)]. Consequently only a lower level value of
W(H$\alpha$) is listed in Table~\ref{pms}.  

\begin{figure}
\includegraphics[width=85mm]{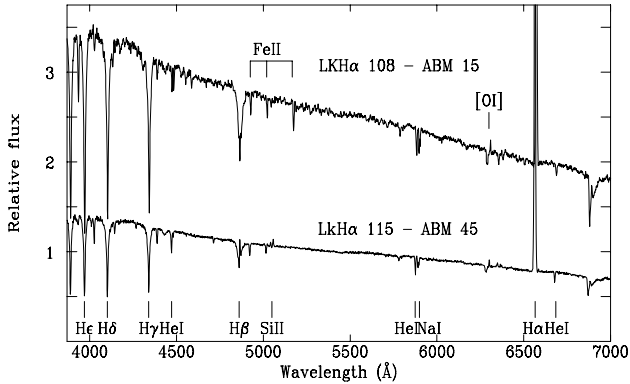}
\caption{Flux calibrated spectra of the two Herbig~Ae/Be stars observed in 
M8. The rest-wavelenghts of the most prominent lines, including the Balmer 
series (H$\epsilon$, H$\delta$, H$\gamma$, H$\beta$, H$\alpha$), the
Fe\,{\sc ii} (42) multiplet, He\,{\sc i} ($\lambda$4471, $\lambda$5876 and 
$\lambda$6678) and Na\,{\sc i}\,$\lambda\lambda$5890, 5896 lines, are indicated.
The high S/N ratio of these spectra allowed us to detect new spectral 
emission features, such as Si\,{\sc ii}\,$\lambda\lambda$5041, 5056 and 
$\lambda\lambda$6347, 6371 and [O\,{\sc i}]\,$\lambda\lambda$6300, 6364, which 
are also marked in the figure.}
\label{HAB}
\end{figure}

Although the original definition of the T~Tauri class includes stars of
spectral types late-F and G, it would be convenient to consider these objects 
intermediate in mass between T~Tauri stars and Herbig Ae/Be stars 
as a different class of PMS objects (Mart\'{\i}n 1997). 
We then adopt the denomination ``PMS~Fe/Ge stars'', suggested by 
Mart\'{\i}n (1997), to refer to this new class of objects.
In column~3 of Table~\ref{pms} we indicate the PMS class for each 
of the observed objects, i.e., classical T~Tauri (CTT), weak T~Tauri (WTT), 
Herbig~Ae/Be (HAeBe) or PMS~Fe/Ge (PMS~FeGe).
The most common criterion used to distinguish between CTT stars and WTT stars 
is based on the equivalent width of the H$\alpha$ emission line 
[W$_\lambda$(H$\alpha$)], 
although the dividing line is still somewhat controversial. The observed value 
of W$_\lambda$(H$\alpha$) also depends on the spectral type. 
Martin (1998) suggests criteria that account for this spectral type 
dependence. More recently, White \& Basri (2003) suggested a slight
modification to these values based on data extracted from the literature for 
a large sample of T~Tauri stars. Specifically, they proposed that a T~Tauri
star is classical if W$_\lambda$(H$\alpha$)\,$\geq 3$\,\AA\, for K0-K5 stars 
and W$_\lambda$(H$\alpha$)\,$\geq 10$\,\AA\, for K7-M2.5 stars.
Thus, according to these empirical boundary values, 27 of the objects in our 
sample are identified as CTT stars, whereas only 7 are classified as 
WTT stars. This was to be expected since most of the targets were selected 
due to their optical colours indicative of strong H$\alpha$ emission.
The three G-type stars are included in the previously defined 
``PMS~Fe/Ge class''.
A sample spectrum of each of the mentioned PMS~groups 
are shown in Figure~\ref{sample}. Figures showing the whole set of observed 
objects are included in an electronically available appendix. 
Finally, the two objects in Figure~\ref{HAB} are LKH$\alpha$~108 (ABM\,15) and
LkH$\alpha$~115 (ABM\,45), classified as Herbig~Be stars. Both stars had been 
identified as part of this stellar class by van den Ancker et al. (1997), who 
also estimated their masses and ages (5.5~$M_{\odot}$ and $10^5$ years for 
both objects). Having resolution and S/N ratio considerably higher than 
previously published data,
our spectra reveal new spectral emission features, such as 
Si\,{\sc ii}\,$\lambda\lambda$5041, 5056 and $\lambda\lambda$6347, 6371 
in ABM\,45 and [O\,{\sc i}]\,$\lambda\lambda$6300, 6364 in both cases, 
and also confirm their membership to the group. 
A few of the observed objects (7) show neither Li\,{\sc i} absorption nor 
emission lines; these are probably field stars not related to the star forming 
region and their spectra have not been included in the figures.

\begin{figure*}
\includegraphics[width=120mm, angle=270]{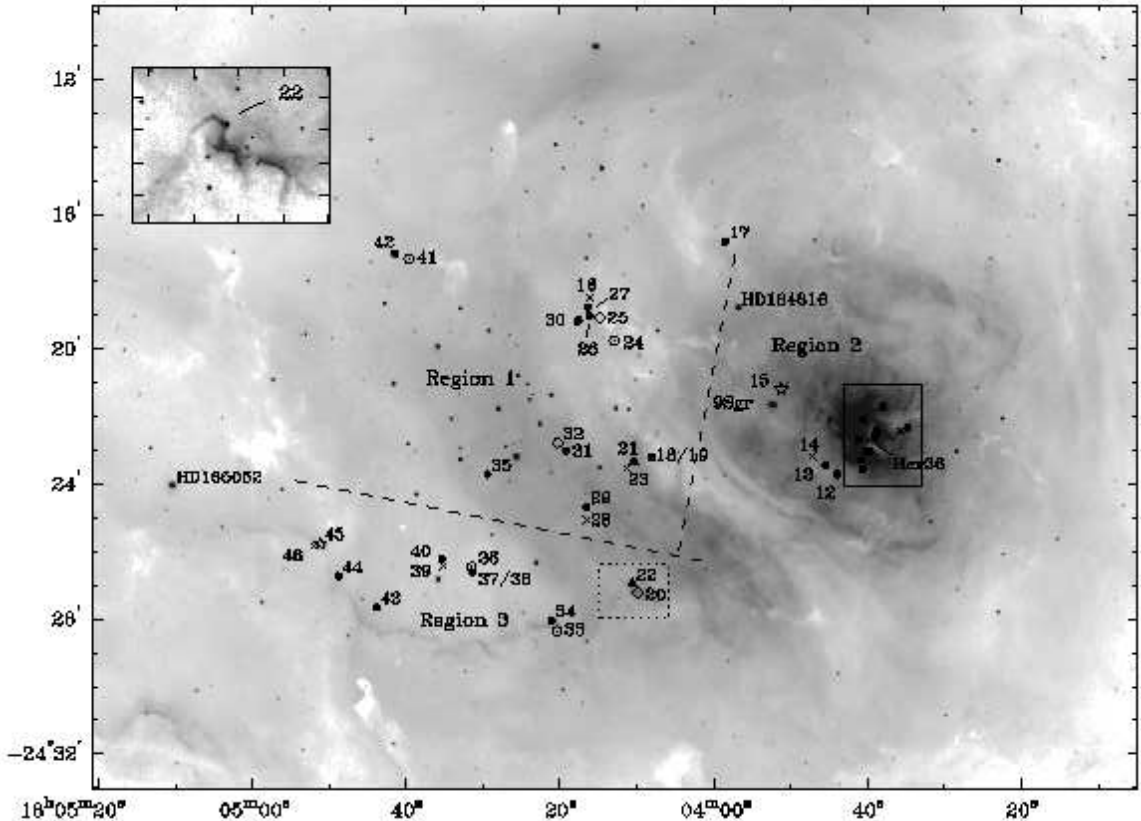}
\caption{The NGC\,6530 field in [S{\sc ii}] obtained with the CTIO 
Curtis-Schmidt telescope showing the complete sample of observed objects. 
CTT (27) and WTT (7) stars are marked by filled and open circles, 
respectively. The filled triangles (3) represent PMS~Fe/Ge stars and the open 
stars (2) indicate the Herbig~Ae/Be objects. The crosses (7) denote some 
observed objects that appear to be normal field stars.
O-type stars in the region (HD~165052, HD~164816, 9\,Sgr and Her\,36) are 
included for reference. An enlargement of the region around ABM\,22
(dot-lined box) showing the finger-like pillar where this object is placed 
(see Sec.~\ref{hh}) is presented in the left-upper corner of the figure.
The solid-lined box (Hourglass nebula) indicates the area shown in 
Figure~\ref{campo_hourglass}.
The dashed lines approximately delimitate the three basic regions discussed 
in section~\ref{HRsec}, which are also labelled as ``Region 1'', ``Region 2'' 
and ``Region 3''.}
\label{campo_m8}
\end{figure*}

\begin{figure}
\includegraphics[width=85mm]{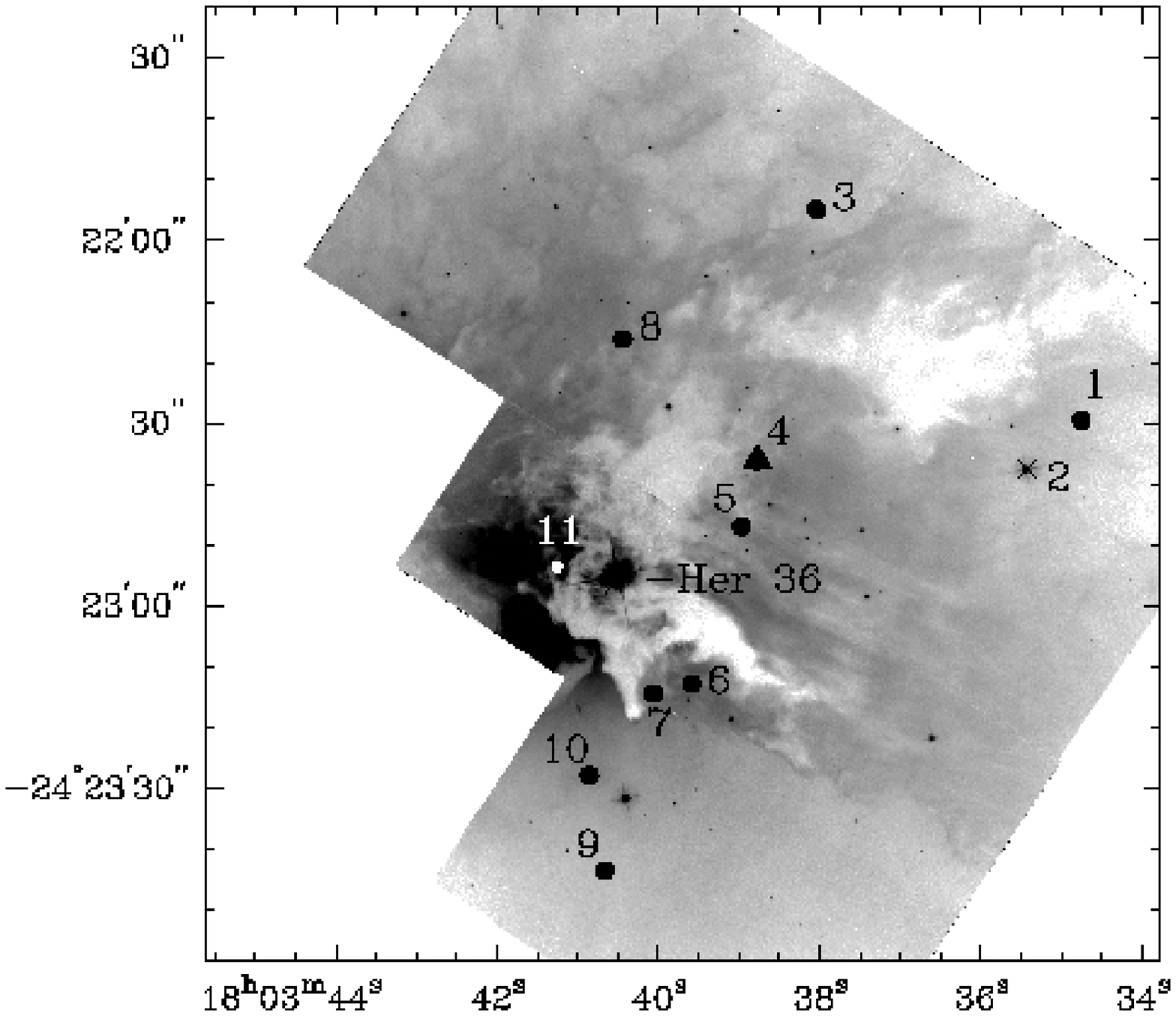}
\caption{The field of the Hourglass nebula observed with 
{\em HST}-WFPC2/F547N, showing the observed objects in the region. Symbols 
are the same as in Figure~\ref{campo_m8}. The O-type star Herschel~36 is also 
shown for reference.}
\label{campo_hourglass}
\end{figure}

Figures~\ref{campo_m8} and \ref{campo_hourglass} show the spatial distribution
of the 46 observed objects in the M8 region. 
Different symbols have been used in order to indicate different object classes.
CTT stars (27) and WTT stars (7) are indicated by filled and open circles,
respectively. The three PMS~Fe/Ge objects are denoted by filled triangles, 
while the two Herbig~Be objects are represented by open stars. 
Finally, the objects marked with crosses (7) lack all the characteristics 
typical of PMS stars and are probably foreground or background objects.  
It must be stressed that only one of the latter objects belonged to 
our original sample of PMS candidates for spectroscopic follow-up 
(ABM\,2 = SCB~106, included in Table~4 of Sung et al. 2000), 
whereas the rest were observed only due to their proximity to other targets.
Thus, we obtained almost a 100\% success rate in the identification of new PMS
stars from the selected targets.
The four O-type stars in the field (HD~165052, HD~164816, 9\,Sgr and Her\,36) 
are included for reference. 
We note here that the new PMS stars are more or less uniformly spread over 
the whole star forming region, with a marked 
concentration of YSOs toward the Hourglass Nebula. 

\begin{figure}
\includegraphics[width=85mm]{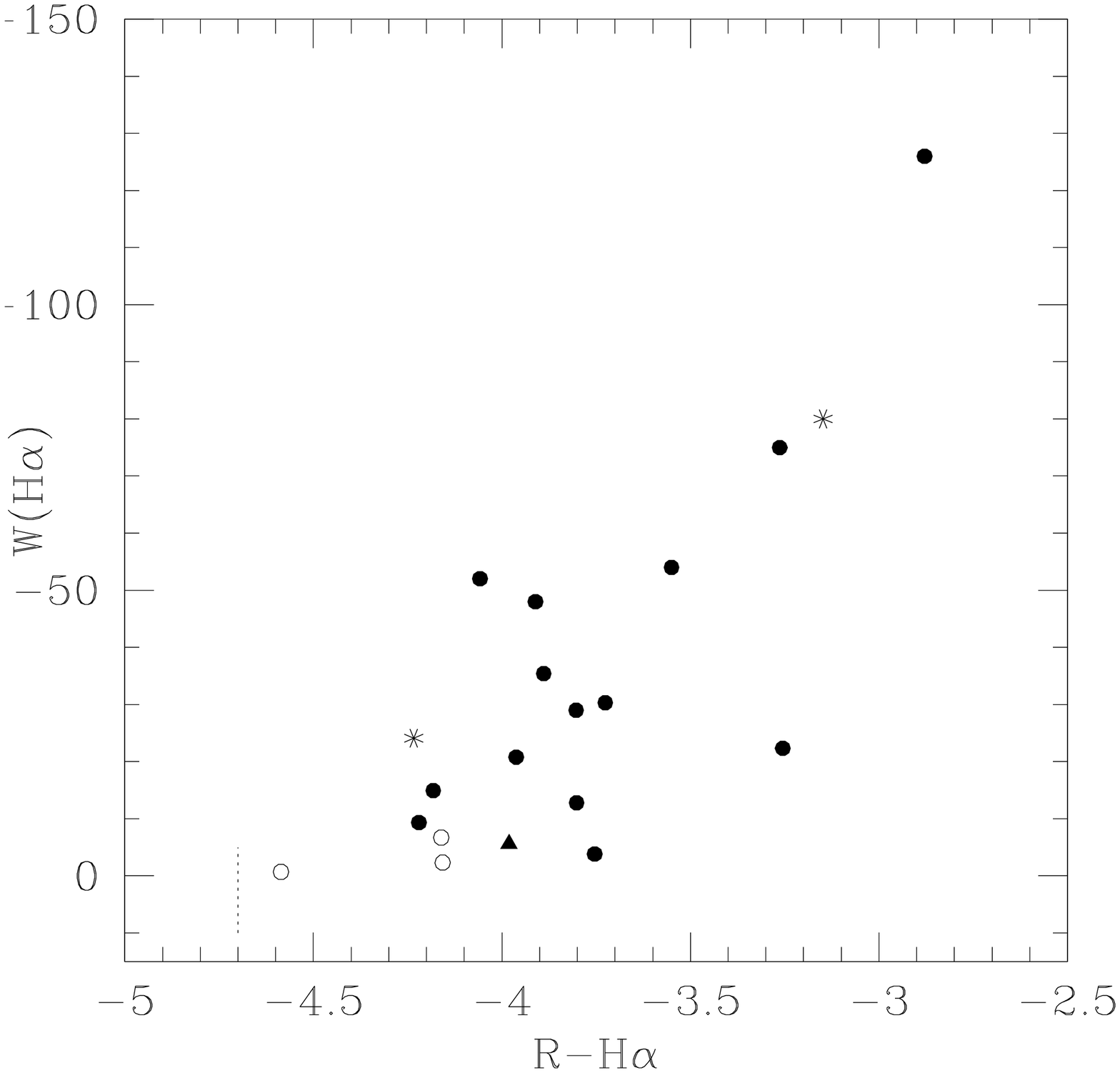}
\caption{Equivalent widths of H$\alpha$ vs. $R-{\rm H}\alpha$ colour from 
Sung et al. (2000) for 21 of the sample objects. The little dashed line 
on the left indicates the average main sequence $(R-{\rm H}\alpha)_{MS}$ 
colour. Symbols are the same as in Figure~\ref{campo_m8}. A reasonably good 
correlation between the two parameters is observed.}
\label{correla}
\end{figure}

Before concluding this section, we briefly discuss whether our measured 
H$\alpha$ equivalent widths agree with what is expected from the 
($R-{\rm H}\alpha$) excess measurements of Sung et al. (2000). 
In Figure~\ref{correla} we have plotted W$_\lambda$(H$\alpha$) against 
($R-{\rm H}\alpha$) for the 21 objects in our sample for which the photometric 
data exist. Symbols are the same as in Figures~\ref{campo_m8} and 
\ref{campo_hourglass}. 
For normal main sequence stars, the ($R-{\rm H}\alpha$) colour varies
relatively little with the ($V-I$) colour (see Figure~6 of Sung et al. 2000). 
Thus one can consider the average main sequence relation 
$(R-{\rm H}\alpha)_{MS}= -4.7 \pm 0.2$, that corresponds to the dashed line 
on the left of Figure~\ref{correla}. Objects with ($R-{\rm H}\alpha$) in excess
of this value must have some kind of H$\alpha$ emsission. As seen in 
Figure~\ref{correla}, there exists a reasonably good correlation between the 
measured H$\alpha$ equivalent width and the ($R-{\rm H}\alpha$) colour. The 
moderate dispersion observed must be accounted for photometric errors
 as well as for intrinsic variability of the H$\alpha$ emission. 
Drew et al. (2005) studied the colours of the emission-line stars.
Using synthetic colours, they analyzed the impact on the 
($r'-{\rm H}\alpha$) vs. ($r'-i'$) colour-colour 
plane of increasingly strong 
H$\alpha$ emission, finding that a given location in this plane,
above the main stellar locus, is associated with a particular H$\alpha$ 
equivalent width. 
We note the same trend in Sung et al. (2000)'s ($R-{\rm H}\alpha$) colour 
with respect to our measured H$\alpha$ equivalent widths.
However we could not perform a further comparison since different filter 
systems have been used in both cases. 


\subsection{Herbig-Haro emission in ABM~22}
\label{hh}

\begin{figure}
\includegraphics[width=48mm, angle=90]{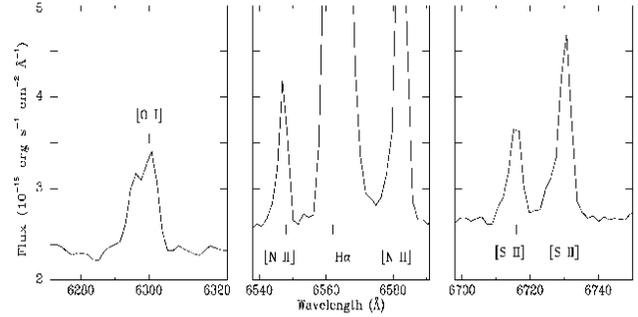}
\caption{Spectrum of ABM\,22 in the regions of the [O\,{\sc I}] $\lambda$6300
  (left-hand panel), H$\alpha$ (middle panel) and [S\,{\sc ii}] 
$\lambda\lambda$6716, 6731 emission lines (right-hand panel), suggesting the
presence of a probable HH~jet associated with this PMS star. The ticks mark the
rest-wavelenghts of the emission features.}
\label{esp22}
\end{figure}

The case of ABM\,22 deserves a special remark. 
Figure~\ref{esp22} shows the spectrum of ABM\,22 in the regions of the 
[O\,{\sc i}] $\lambda$6300 forbidden line (left-hand panel), 
the H$\alpha$ and [N\,{\sc ii}]\,$\lambda\lambda$6548, 6584 lines 
(middle panel) and the [S\,{\sc ii}]\,$\lambda\lambda$6716, 6731 lines
(right-hand panel).
Among other young stellar objects (YSOs), optically bright T~Tauri stars, and
certainly also the more massive PMS~Fe/Ge objects, are 
known to be sources of Herbig-Haro (HH) jets. 
As seen in Fig.~\ref{esp22}, the H$\alpha$ and the 
[N\,{\sc ii}]\,$\lambda$6583 lines  in the spectrum of ABM\,22 are 
particularly strong, exhibiting emission blueshifted to relatively high 
velocities.  The same  behavior is observed in the [S\,{\sc ii}] lines.
While this blueshifted
emission is barely visible for the H$\alpha$ and [N\,{\sc ii}] lines 
as it clearly runs into the adjacent features, the profiles of the 
[S\,{\sc ii}] lines suggest velocities of at least 300 kms$^{-1}$. 
It has been argued that the high-velocity forbidden line emission of these PMS
stars is produced by a well-collimated jet (Kwan \& Tademaru 1988), an idea 
that is supported by several observational studies (e.g. Hirth, Mundt, \& Solf
1994, 1997).
More remarkably, the spectrum of ABM\,22 shows strongly double peaked
emission lines in the forbidden [O\,{\sc i}]~$\lambda\lambda$~6300, 6363 
doublet (for clarity, only the [O\,{\sc i}] $\lambda$ 6300 line is shown in
Fig.~\ref{esp22}).
The widths of the [O\,{\sc i}] line profiles are unusually large,  
nearly twice the width of an ordinary nebular line, which 
suggests that they are actually double lined features, composed by a 
redshifted and a blueshifted component, presumably originated in the receding 
and approaching flows, respectively.

Mundt \& Eisl\"offel (1998) have shown that T~Tauri stars that exhibit 
spectroscopic 
evidence for mass loss, such as strong forbidden lines, have a high 
probability of being associated with an extended jet or HH emission.
Based on narrowband images obtained with ESO 2p2/WFI and retrieved from the 
ESO public data base, 
we have recently identified a number of nebular structures 
distributed all over the field of the M8 nebula, which appear as prime 
candidates to be small- and large-scale HH outflows (Barb\'a \& Arias 2006). 
One of them seems to be related to the source ABM\,22.
ABM\,22 is in fact located in the tip of a finger-like dusty pillar which 
points toward the massive star 9\,Sgr (see the enlargement on the left-upper
corner of Figure~\ref{campo_m8}). 
A nebular feature, very bright in [S\,{\sc ii}], is aligned with the star,
stretching for about $3'$ from the end of the dust finger.  This
feature closely resembles the outflows originating in the molecular cloud core 
that are breaking into the surrounding H\,{\sc ii} region reported by Bally 
et al. (2002) for S140. 

Although the observational evidence presented in this paper strongly suggests 
the existence of a HH object associated with ABM\,22, we are reluctant to draw 
any conclusion here, especially because of the 
relatively low resolution of our spectroscopic observations.
A future high-resolution study will be needed to confirm this possibility.

\begin{figure}
\includegraphics[width=53mm, angle=90]{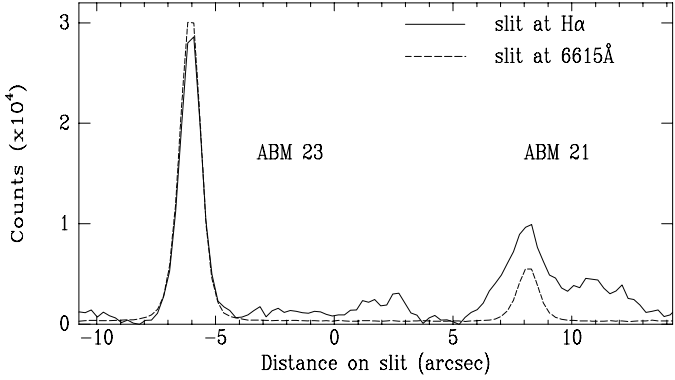}
\caption{Cuts of the slit along the position of the H$\alpha$ line (solid line)
and the continuum at 6615\,\AA\,(dashed line). In order to compare the two
spectra, a quadratic fit of the background was subtracted to both of them.}
\label{abm21}
\end{figure}

To conclude this section, we make here a short comment about three particular
objects in the centre of NGC~6530: ABM\,21, ABM\,27  and ABM\,29. 
Their spectra show unexpected [O\,{\sc iii}]\,$\lambda\lambda$4959, 5007 
emission lines, which furthermore are conspicuously strong in all cases.
After discarding an inaccurate background subtraction, 
we examined the stellar profiles along the slit. 
We found that the stellar profiles in the [O\,{\sc iii}],   
H$\alpha$ and H$\beta$ emission lines in the spectrum of ABM\,21 
are much wider (FWHM $\sim 1\farcs6$) than
the stellar continuum emission profile (FWHM $\sim1''$),
suggesting  that this star is located in a bright high-excitation knot.
This can be appreciated in Figure~\ref{abm21} which shows two cuts of the slit 
along the position of the H$\alpha$ line (solid line) and 
the continuum at 6615\,\AA\,(dashed line), respectively.
The other star on the slit is the foreground object ABM\,23, for which 
the two stellar profiles coincide since it shows no H$\alpha$ emission.
On the other hand, the H$\alpha$ emission of ABM\,21 is remarkably stronger
and broader with respect to the continuum emission. An emission bump $5''$
wide on which the star seems to be located is also noticeable.
ESO 2p2/WFI H$\alpha$ images 
also confirm the extended emission from this source. 
Although somewhat less evident, the cases of ABM\,27 and ABM\,29 are very 
similar and  these stars would also be located in a high-excitation knot.
Again, we expect that future high-resolution observations will help to reveal 
the nature of these interesting objects.

\subsection{Near-infrared properties}

Using data from 2MASS and from our own previous 
infrared photometry (Arias et al. 2006), we constructed the {\em JHK$_s$} 
colour-colour (CC) and colour-magnitude (CM) diagrams for the observed objects 
(Figure~\ref{diagIR}).
Also shown in the CC diagram are the position of the main sequence and 
red giant stars and the locii of T~Tauri stars, indicated as a band 
defined by the track from Meyer et al. (1997) and a hotter track roughly
corresponding to a K0 star temperature. 
The parallel line below the TTS~locii labelled as ``Fe/Ge locus'' represents 
the expected positions for PMS Fe/Ge stars, i.e. PMS objects
with spectral types late-F and G, as introduced in section~\ref{clasifica}. 
The two parallel dashed lines correspond to the reddening vectors for 
early- and late-type stars and define the reddening band. 
The reddening vector for a K0 dwarf is also indicated for reference, as a 
dotted line within this band. 
In the CM diagram, the position of the main sequence has been plotted, 
corrected to an apparent distance modulus of 10.5, which seems appropiated 
for NGC\,6530 according to the last determinations by Prisinzano et al. (2005)
and Arias et al. (2006).
The reddening vector for an O7\,V is also plotted for a visual extinction
of 20 mag. CTT and WTT stars are denoted by filled and open circles, 
respectively. The filled triangles represent PMS~Fe/Ge stars 
whereas the asterisks indicate the Herbig~Ae/Be objects. 
The observed foreground objects are marked with 
crosses\footnote{The stars ABM\,2 and ABM\,28 have been omitted in the diagrams
  because of the low quality of their 2MASS photometry. The inspection of
 the 2MASS $K_s$-band image shows that ABM\,2 has a bright  infrared companion 
 1\farcs9 distant which has contaminated its magnitudes. The $K_s$-magnitude 
 of ABM\,28 must be spurious, as evidenced from the comparison with the
 $K_s$-magnitudes of neighboring infrared sources in the field.}.

The positions of the newly discovered late-type PMS stars in the CC~diagram 
are compatible with those expected for T~Tauri stars affected by low to 
moderate reddening (up to about 7 mag), with the exception of ABM\,1, 
the only star with negative $H-K_s$. The location of this faint star strongly 
deviates from the group, which can presumably be explained by the low quality 
of its 2MASS photometry ($ph\_qual={\rm CDB}$\footnote{The photometric 
quality flag $ph\_qual$ provides a guide to the quality of the default 
point source photometry that is based on signal to noise ratio, measurement 
quality, detection statistics, etc. It is comprised of three characters, each 
corresponding to one band. Sources with $ph\_qual={\rm A}$ in a band, 
have $<10\%$ measurement uncertainties. Measurement quality decreases with 
alphabetically increasing values of $ph\_qual$.}). 
Most of the targets do not present substantial infrared excesses.
However some of them (12) are located outside the reddening band, more than 
0.25 mag (in $H-K_s$) to the right of the K0 reddening vector, 
showing significant excess emission. 
The two observed Herbig Be stars also present important 
near-infrared excesses. 
As expected for PMS stars, our objects are overluminous with respect to 
main sequence stars with the same spectral types. This can be clearly 
appreciated in the CM diagram where they show a {\em K$_s$}-band magnitude 
which is 3-4 mag brigther than the corresponding to a typical K-type dwarf.

\subsection {X-ray emission}
\label{XR}
 
It is a well known fact that PMS stars usually have strong X-ray emission. 
We thus positionally matched our observed stars with the X-ray sources 
detected with the {\em Chandra} ACIS instrument, recently published by 
Damiani, Flaccomio, Micela et al. (2004). 
The cross-correlation between our and their sources (with nomenclature 
``DFM2004'') is shown in the tenth column of Table~\ref{literature}.
We found that all but two of the newly discovered late-type PMS stars 
contained in the FOV of the {\em Chandra} observation, including 
both {\em classical} and {\em weak-lined} T~Tauri stars and 
PMS~Fe/Ge objects, present X-ray emission.
Unfortunately the Hourglass region, which harbors an important fraction of 
the new PMS objects, could not be observed because of the limited FOV of the
instrument. 
As was to be expected, the two Herbig~Be objects in our sample are 
also identified as strong X-ray emitters.
The count rates measured for all these X-ray sources are listed in the last 
column of Table~\ref{pms}.

\begin{table*}
\begin{minipage}{180mm}
\caption{Spectroscopy and stellar properties of the observed objects in M8. Running source numbers are in column 1. Spectral types and PMS~classes are in columns 2 and 3, respectively. 
Columns 4 and 5 contain the measured equivalent widths of the H$\alpha$ and the Li\,{\sc i} lines. Additional emission lines observed in the spectra are indicated in column 6. Bolometric 
luminosities and effective temperatures derived in Sec.~\ref{discu_1} are in columns 7 and 8. Finally, the X-count rates (X-CR) expressed in counts~ks$^{-1}$ are listed in column 9.}
\label{pms}
\begin{tabular}{ccccclccc}
\hline\\
Object    &  Sp.T.        & Type  &  W(H$\alpha$) & W(Li\,{\sc i})  &  Other emission lines                           & log($L_{bol}/L_{\odot}$) & log($T_{eff}$) & X-CR$^\ast$\\
ABM       &               &       &    [\AA]      & [\AA]           &                                                 &               &                                & [c\,ks$^{-1}$] \\
$[1]$ &  [2]            &   [3]   &     [4]       &    [5]          &  ~~~~~~~~~~~[6]                              &    [7]           &    [8]           & [9]        \\
\hline \\
1         & K2-3\,V       & CTT   & $<$-23. (2.)  &  0.85 (0.07)    & Balmer lines (BL),  [S\,{\sc ii}]     & -0.37 & 3.682 &\\
2         & G2\,IV-V      & ...   &  2.78 (0.07)  &   ...           & ...                                                                                     & & &\\
3         & K5-7\,V       & CTT   & $<$-44. (5.)  &  0.58 (0.05)    & Ca\,{\sc ii}, BL, He\,{\sc i}, [O\,{\sc i}], [S\,{\sc ii}]                 & 0.22 & 3.624 &\\
4         & G0-8          & PMS~Ge   & $<$-95. (12.) &  0.26 (0.05) & BL, He\,{\sc i}, [S\,{\sc ii}], [N\,{\sc ii}], [O\,{\sc i}]                                 & 0.73 & 3.761 &\\
5         & K7\,V         & CTT   & -220. (4.)    &  0.79 (0.04)    & BL, He\,{\sc i}, [O\,{\sc i}], [S\,{\sc ii}]                                                & 0.85 & 3.608 &\\
6         & K3\,V         & CTT   & -264. (48.)   &  0.56 (0.08)    & Ca\,{\sc ii}, BL, He\,{\sc i}, Fe\,{\sc ii}, [O\,{\sc i}], [N\,{\sc ii}], [S\,{\sc ii}]     & 0.84 & 3.675 &\\
7         & K7\,V         & CTT   & -94. (5.)     &  1.04 (0.04)    & BL, He\,{\sc i},  [S\,{\sc ii}]                                                              & 0.51 & 3.608 &\\
8         & K7\,V         & CTT   & $<$-76. (3.)  &  0.77 (0.09)    & H$\beta$, He\,{\sc i}, [O\,{\sc i}],  [S\,{\sc ii}]                                          & 1.07 & 3.608 &\\
9         & K5\,V         & CTT   & -39.0 (0.6)   &  0.82 (0.07)    & Ca\,{\sc ii}, BL, [N\,{\sc ii}], [S\,{\sc ii}]                                & 0.81 & 3.638 &\\
10        & K4-5\,V       & CTT   & -29.8 (0.9)   &  0.46 (0.04)    & Ca\,{\sc ii}, BL, He\,{\sc i}, [S\,{\sc ii}], [N\,{\sc ii}]                   & 1.24 & 3.650 &\\
11        & K0-2\,V       & CTT   & -68. (2.)     &  0.45 (0.02)    & BL, He\,{\sc i}, [N\,{\sc ii}], [S\,{\sc ii}]           & 1.19 & 3.705 &\\ 
12        & K5\,V         & CTT   & -15.1 (0.1)   &  0.44 (0.04)    & Ca\,{\sc ii}, BL, He\,{\sc i}, [N\,{\sc ii}]                                 & 0.51 & 3.638 &\\
13        & K4-5\,V       & CTT   & -126. (9.)    &  0.25 (0.02)    & Ca\,{\sc ii}, BL, He\,{\sc i}, Fe\,{\sc ii}, [Fe\,{\sc ii}], [S\,{\sc ii}], [N\,{\sc ii}], & 0.83 & 3.650 &\\
          &               &       &               &                 &  [O\,{\sc i}], Na\,{\sc i}                                   & & \\
14        &  F8\,V        & ...   &  3.27 (0.07)  &   ...           & ...                                                                                     & & &\\
15        &  B6\,Ve       & HAeBe & -24.1 (0.4)   &   ...           & H$\beta$, [O\,{\sc i}]                                                                  & & & 0.769\\
16        &  B8-9\,V      & ...   &  13.7 (0.9)   &   ...           & ...                                                                                     & & &\\
17        &  K4-5\,V      & CTT   & -52. (3.)     &  0.25 (0.02)    & Ca\,{\sc ii}, BL, Fe\,{\sc ii}, [Fe\,{\sc ii}], Ti\,{\sc ii}, He\,{\sc i}, Na\,{\sc i} & 0.32 & 3.650 & 0.890 \\
18        &  K5\,V        & CTT   &  -3.8 (0.2)   &  0.60 (0.01)    & ...                                                                          & 0.19 & 3.638 & 1.742\\
19        &  K2\,V        & CTT   & -16.0 (0.6)   &  0.57 (0.03)    & Ca\,{\sc ii}, H$\beta$, [N\,{\sc ii}], [S\,{\sc ii}]                                 & 0.64 & 3.690 & 1.742\\
20        &  K5\,V        & WTT   &   2.7 (0.1)   &  0.46 (0.04)    & Ca\,{\sc ii}                                                                         & 0.36 & 3.638 & 0.901\\
21        &  G0-8\,V        & PMS~Ge   &  -5.61 (0.03) &  ...         & BL, [N\,{\sc ii}], [O\,{\sc iii}]                                                  & 0.97 & 3.761 & 0.212\\
22        &  G0-8\,V     & PMS~Ge   &  -27.8 (0.5)  &  0.10 (0.01)    & Ca\,{\sc ii}, Fe\,{\sc ii}, BL, [O\,{\sc i}], [N\,{\sc ii}], [S\,{\sc ii}]    & 1.33 & 3.761 & 0.267\\
23        &  K5\,III      & ...   &   1.0 (0.1)   &  ...            & ...                                                                                     & & &\\
24        &  K1-2\,V      & WTT   &  -0.65 (0.03) &  0.49 (0.03)    & ...                                                                          & 0.85 & 3.698 & 4.420\\
25        &  K7\,V        & WTT   & -8.1 (0.30)   &  0.70 (0.01)    & BL,  Ca\,{\sc ii}                                                            & 0.09 & 3.608 & 0.881\\
26        &  K4\,V        & CTT   &  -9.3 (0.2)   &  0.44 (0.03)    &  Ca\,{\sc ii}                                                           & 0.57 & 3.661 & 0.744\\
27        &  K0-3         & CTT   & -75. (4.)     &  0.26 (0.01)    & Ca\,{\sc ii}, BL, Fe\,{\sc ii}, He\,{\sc i}, [O\,{\sc iii}], [O\,{\sc i}], [N\,{\sc ii}]      & 0.51 & 3.698 & 0.398\\
28        &  F7\,V        & ...   &   4.75 (0.15) &  ...            & ...                                                                                     & & &\\
29        &  K4\,V        & CTT   &  -29. (1.)    &  0.98 (0.05)    & Ca\,{\sc ii}, BL,  [O\,{\sc iii}], He\,{\sc i}, [N\,{\sc ii}]                & 0.51 & 3.661 & 2.373\\
30        &  K5\,III      & CTT   &  -30.3 (0.7)  &  0.34 (0.04)    & Ca\,{\sc ii}, BL, He\,{\sc i}, [O\,{\sc i}], [N\,{\sc ii}], [S\,{\sc ii}]     & 1.08 & 3.638 & 1.242\\
31        & K4\,III       & CTT   & -14.96 (0.04) &  0.94 (0.06)    & Ca\,{\sc ii}, BL, He\,{\sc i}, [N\,{\sc ii}], [S\,{\sc ii}]                   & 0.76 & 3.662 & 1.614\\
32        & K5\,III       &  WTT  &  -6.7 (0.4)   &  0.64 (0.03)    &  Ca\,{\sc ii}, BL                                                             & 0.29 & 3.638 & 2.424\\
33        & K2-3\,III     &  WTT  &  -2.26 (0.07) &  0.95 (0.07)    & Ca\,{\sc ii}, BL, [N\,{\sc ii}], [S\,{\sc ii}]                                & 1.16 & 3.683 & 1.820\\
34        & K4-5          &  CTT  & -12.8 (0.3)   &  0.54 (0.02)    &  BL, [N\,{\sc ii}], [S\,{\sc ii}]                                             & 0.801 & 3.650 & 1.579\\
35        & K3-4\,V       &  CTT  & -35.4 (0.7)   &  0.72 (0.06)    & Ca\,{\sc ii}, BL, He\,{\sc i}                                                 & 0.43 & 3.668 & 0.700\\
36        & K3-4\,V       &  WTT  &  -0.70 (0.07) &  0.41 (0.05)    & Ca\,{\sc ii}                                                                  & 0.67 & 3.668 & 19.963\\
37        & K5\,V         & CTT   & -30.3 (0.5)   &  0.34 (0.04)    &  Ca\,{\sc ii}, BL, He\,{\sc i},  [O\,{\sc i}]                        & 0.41$^{\dag}$ & 3.638 & 5.478\\
38        & K5            & CTT   & -20.4 (0.2)   &  0.89 (0.03)    & Ca\,{\sc ii}, H$\beta$,  He\,{\sc i}                                                       & & &\\
39        & K3-4\,III     & ...   &  1.45 (0.09)  &  ...            & ...                                                                                     & & &\\
40        &  K4\,V        & CTT   & -48. (2.)     &  0.49 (0.04)    & Ca\,{\sc ii}, BL, He\,{\sc i}, Fe\,{\sc ii}                                   & 0.23 & 3.662 &\\
41        & K7\,V         & WTT   & -3.18 (0.06) & 0.80 (0.09)   &  Ca\,{\sc ii}, H$\beta$                                                            & 0.96 & 3.608 & 0.350\\
42        & K4-5\,V       & CTT   & -54. (2.)     &  0.83 (0.05)    & Ca\,{\sc ii}, BL, He\,{\sc i}, [S\,{\sc ii}], [O\,{\sc i}]                    & 0.60 & 3.650 & 0.858\\
43        & K4\,V         & CTT   & -20.8 (0.3)   &  0.56 (0.02)    & Ca\,{\sc ii}, BL, [O\,{\sc i}], He\,{\sc i}, [N\,{\sc ii}], [S\,{\sc ii}]     & 1.11 & 3.662 & 1.215\\
44        &  K5\,V        & CTT   & -22.3 (0.4)   &  0.70 (0.07)    & Ca\,{\sc ii}, BL, He\,{\sc i}                                                 & 1.26 & 3.638 &\\
45        &  B2\,Ve       & HAeBe & -80. (1.)     &  ...            &  H$\beta$, Si\,{\sc ii},  [O\,{\sc i}]                  & & & 12.377\\
46        &  K2\,V        & ...   &  -2.85 (0.2)  & ...             & ...  
& & &\\
          &               &       &               &                 &                                                                                         & & &\\
\hline\\
\end{tabular}
\footnotesize 
$^\ast$ Values from Damiani, Flaccomio, Micela et al. (2004).\\
$^\dag$ Being ABM\,37 and ABM\,38 the components of a single 2MASS source,
the value of $L_{bol}$ estimated from the infrared magnitudes has been assigned to the brightest object.\\
\end{minipage}
\end{table*}

\section{Discussion}

\subsection{Effective temperatures and bolometric luminosities}
\label{discu_1}

\begin{figure*}
\begin{minipage}{180mm}
\includegraphics[width=90mm]{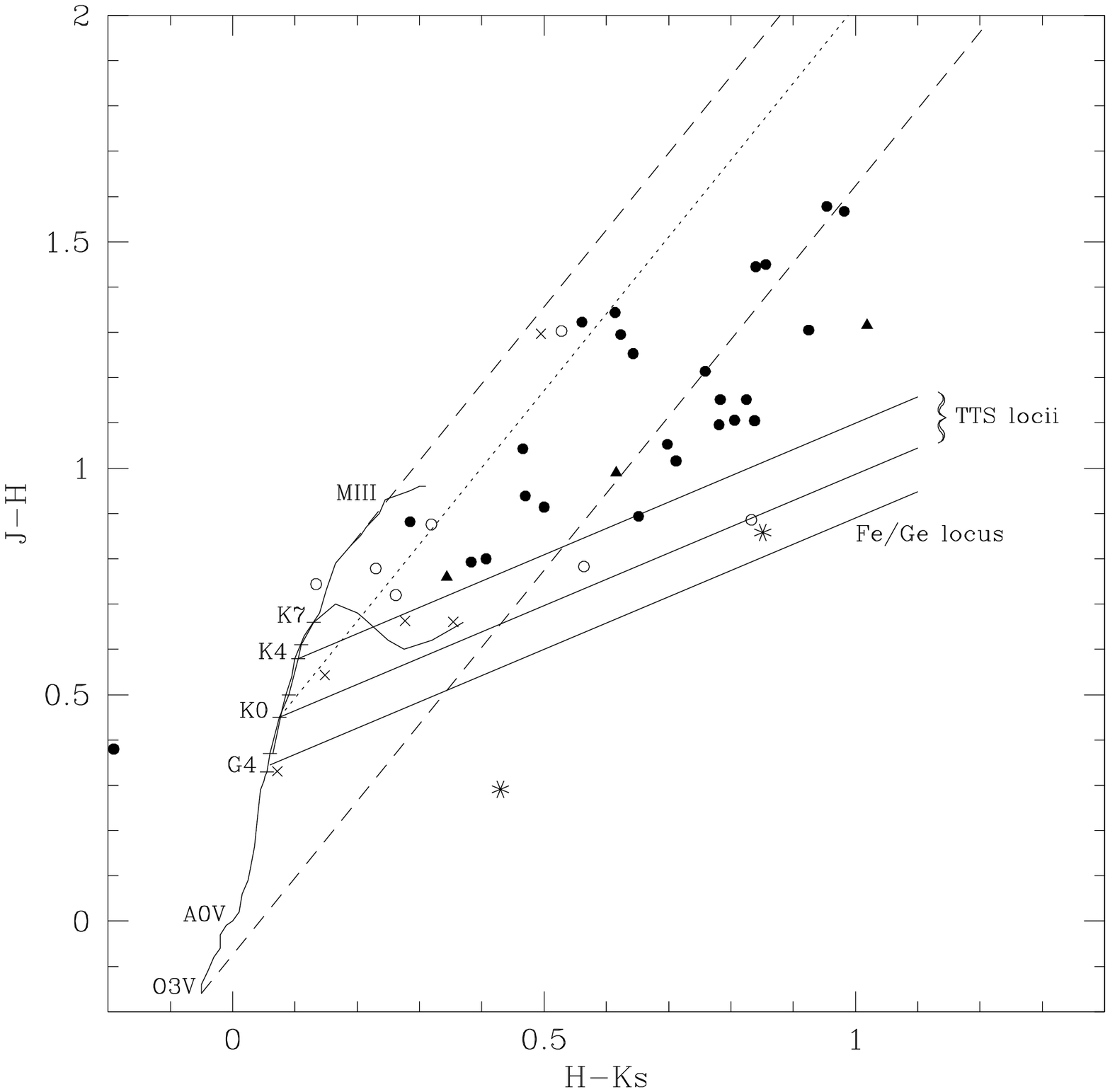}
\includegraphics[width=90mm]{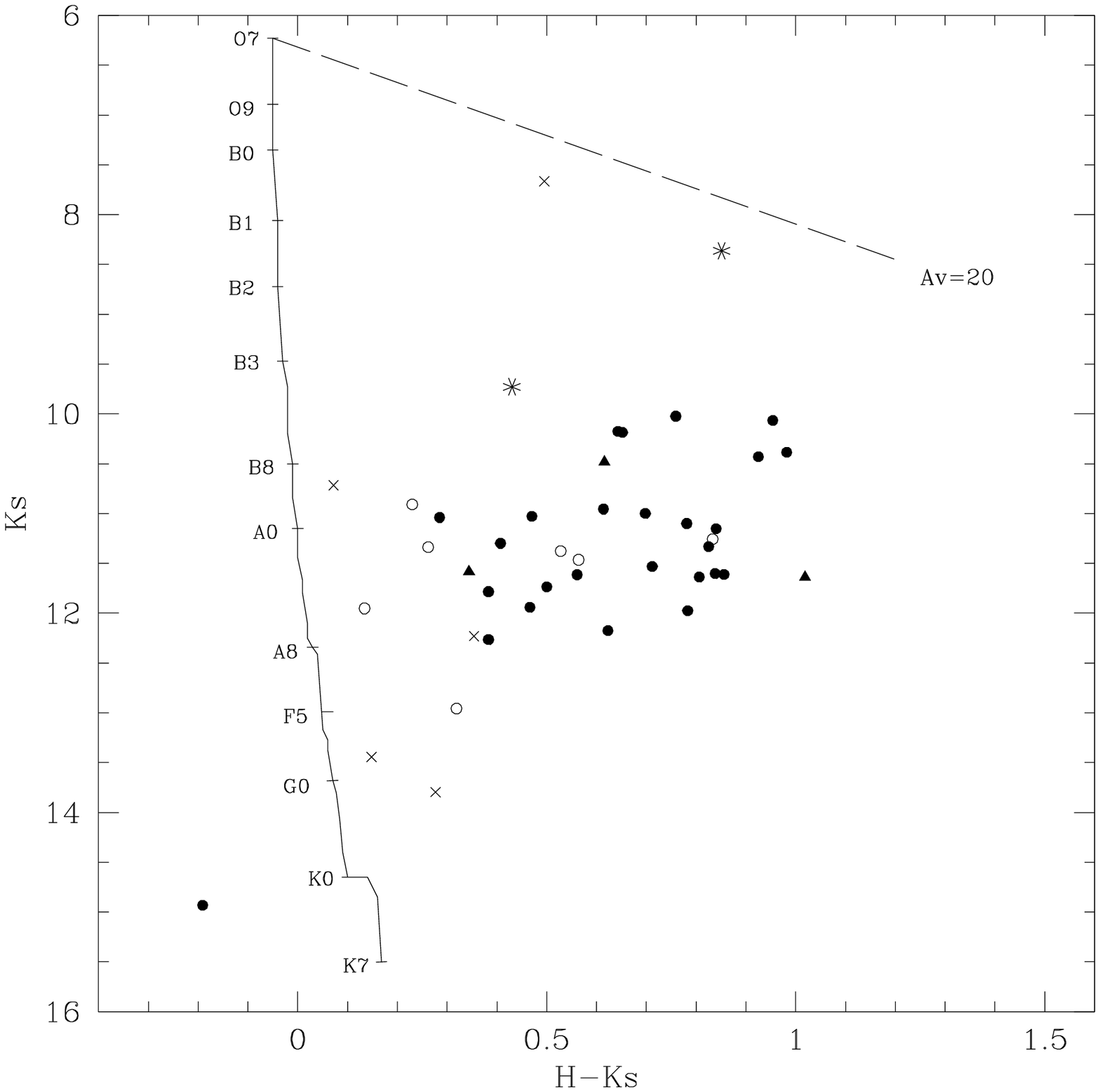}
\end{minipage}
\caption{Left-hand panel: {\em JHK$_s$} colour-colour diagram for the observed 
objects. The loci of the main sequence, giant branch, T~Tauri stars and 
PMS~Fe/Ge objects, as well as the slope of the interestellar reddening, are 
indicated. CTT and WTT stars are denoted by filled and open circles, 
respectively. The filled triangles represent PMS~Fe/Ge stars and the asterisks 
indicate the Herbig~Ae/Be objects. The crosses refer to some observed 
foreground objects.
Right-hand panel: {\em JHK$_s$} colour-magnitude diagram for the observed
objects. The main sequence for a distance modulus of 10.5 mag and the standard
reddening vector with length $A_V=20$~mag are shown. 
Symbols are the same as in the CC~diagram.}
\label{diagIR}
\end{figure*}

We used {\em J}, {\em H} and {\em K$_s$} magnitudes either from 2MASS 
or from Arias et al. (2006) to place the target stars in the
Hertzsprung-Russell (HR) diagram. For this purpose, we had to derive
effective temperatures ($T_{eff}$) and bolometric luminosities ($L_{bol}$).
$T_{eff}$ for each star was obtained from its spectral type according to
Kenyon \& Hartmann (1995). On the other hand, $L_{bol}$ was determined from
the near-infrared photometric data following in part the same procedure as 
described by Kun et al. (2004). The positions of our objects in the 
{\em JHK$_s$} colour-colour diagram are shown in the left-hand panel of 
Figure~\ref{diagIR}. Making the widely used assumption that the total emission
of the star in the {\em J} band originates in the photosphere
(Hartigan et al. 1994), the colour index $J-H$ can be written as

\begin{center}
$J-H = (J-H)_0 + E_{CS}(J-H) + E_{IS}(J-H)$,\\
\end{center}

\noindent where $(J-H)_0$ is the true photospheric colour of the star,
$E_{CS}(J-H)$ is the colour excess in the $H$ band due to the emission of the
circumstellar disk and $E_{IS}(J-H)$ is the colour excess due to the
difference of interstellar extinction in the $J$ and $H$ bands.
In order to determine $E_{IS}(J-H)$, we dereddened our objects 
onto the appropriate PMS stars locii in the colour-colour diagram. 
We considered the locus of classical T~Tauri stars defined by Meyer et al. 
(1997) for objects with types K4 or later.
As the latter is likely too cold for earlier objects, we used instead another 
locus, roughly corresponding to a K0 star temperature, to deredden the objects 
with types K0-K3.  Both tracks are marked in Figure~\ref{diagIR}, 
defining a kind of T~Tauri stars band. 
We note here that the same dereddening proceedure was applied to the WTT 
stars. According to their location in the CC diagram,
most of them are not affected by significant circumstellar extinction
as expected for this class of PMS stars.
A colour excess $E_{IS}(J-H)=0$ was assumed for the two T~Tauri stars located 
below the concerning locii.  
Among our sample of PMS stars, there are also three objects classified as G, 
which 
consequently belong to the PMS~Fe/Ge class defined in section~\ref{clasifica}.
PMS~Fe/Ge stars occupy a different region in the $JHK_s$ diagram, clearly 
below the T~Tauri stars locii, as demonstrated by Hern\'andez et al. (2005).  
Considering the previous locii for them 
would necessarily lead to an underestimation of the extinction.
Thus, these three objects were dereddend onto an approximate locus of 
PMS~Fe/Ge objects (Fe/Ge~locus in Fig.~\ref{diagIR}), 
chosen to  correspond to a middle G star temperature.
Finally, once known the colour excess for every object, 
using the interestellar extinction law $A_J=2.65 \times E_{IS}(J-H)$ 
(Rieke \& Lebofsky 1985) and the bolometric corrections for the $J$ band 
tabulated by Hartigan et al. (1994) we derived the bolometric luminosities
for all the targets.

\subsection{Hertzsprung-Russell diagram}
\label{HRsec}

\begin{figure*}
\includegraphics[width=170mm]{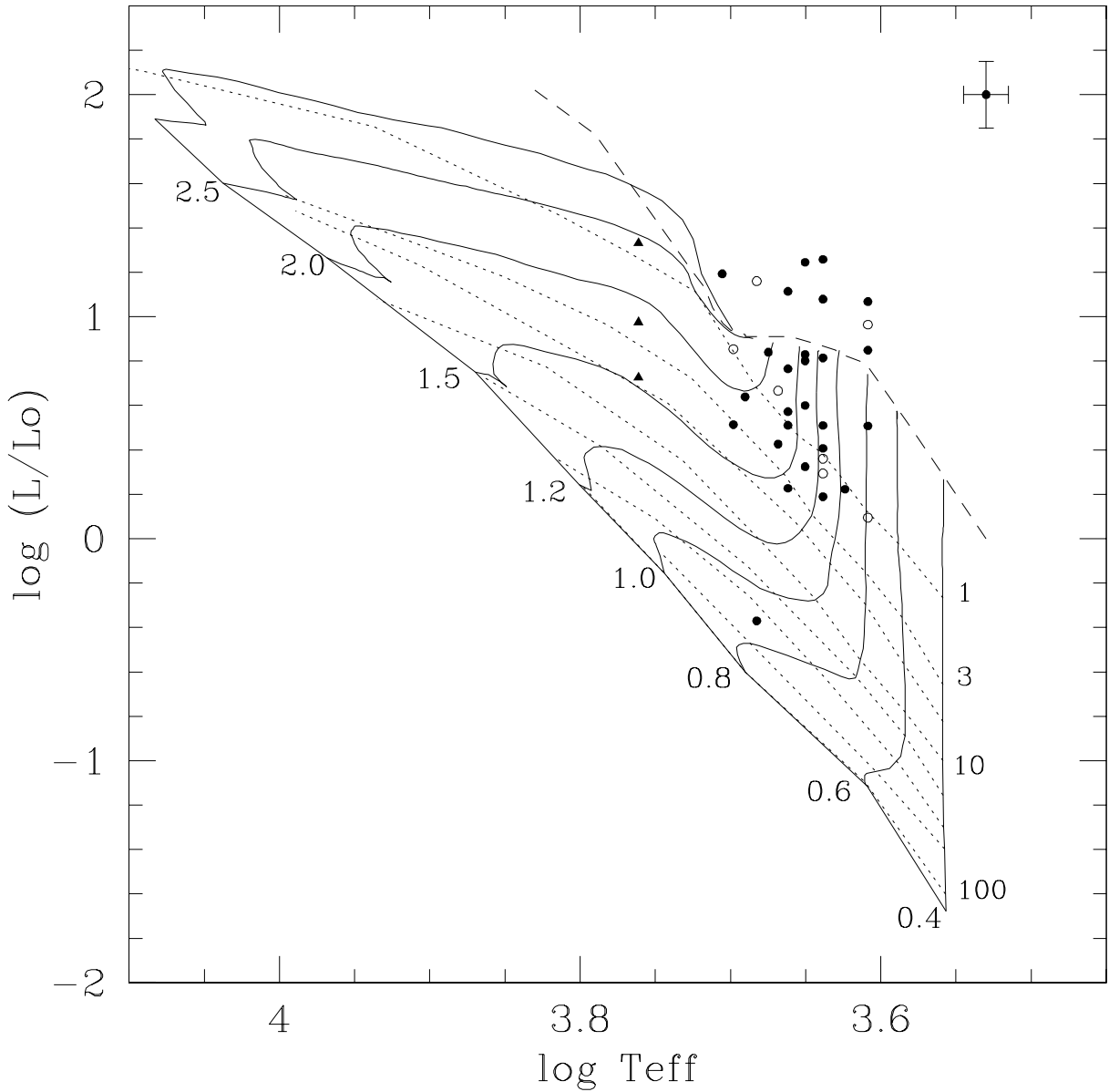}
\caption{Positions of the observed stars in the HR diagram, assuming a 
distance modulus of 10.5. Dotted lines represent the isochrones of 1, 3, 5, 
10, 20, 30, 50 and 100 Myrs and solid lines indicate the evolutionary tracks 
for star masses between 0.4 and 3.0~$M_\odot$ (Palla \& Stahler 1999).
The birthline (dashed line) and the zero age main sequence are also shown.
The location of the very faint star ABM\,1 on the 50~Myr isochrone is 
probably spurious, given the low quality of its 2MASS photometry.}
\label{HRD}
\end{figure*}

The location of the observed PMS stars in the HR diagram is shown in 
Figure~\ref{HRD}. A distance modulus of 10.5 has been assumed for all the 
objects. The filled and open circles denote CTT and WTT stars, respectively,
whereas the filled triangles represent the three PMS~Fe/Ge objects. 
Note that {\em classical} and {\em weak-lined} T~Tauri stars
appear to be mixed in the same region of the HR diagram, which 
corresponds to totally convective Hayashi tracks.
Evolutionary tracks and isochrones, as well as the birthline and
zero age main sequence loci, are also indicated in the figure 
(Palla \& Stahler 1999). The horizontal error bar on the filled circle in the
upper right corner of the figure indicates the 
shift that the objects would experiment due to a $\pm1$ subclass 
uncertainty in the spectral classification 
($\Delta T \approx 160 K$ or 0.015 in logarithmic scale). Similarly,
the vertical error bar represents a ``typical'' uncertainty in the estimated
luminosities.
Luminosity determinations have generally relatively large uncertainties
(typically a factor of 2 or $\sim$ 0.3 in logarithmic scale). Although the 
effect of this on mass determinations is not too relevant for young stars 
with spectral types later than K3, for which the evolutionary tracks are 
almost vertical, it might be important for the earlier objects.  
Additional uncertainties in the bolometric luminosity come from the 
hypothesis that the $J$-band flux is not contaminated by the nonphotospheric
emission of the star. CTT stars present significant excess in all three 
near-infrared bands.  
Cieza et al. (2005) demonstrated that deriving their stellar luminosities by 
applying the ``standard'' procedure, i.e. making bolometric corrections to 
the $J$-band fluxes, systematically overestimates them.
They found that the luminosities derived from the $J$-band would be 
higher by a factor of $\sim1.35$ on average ($\sim$ 0.13 in logaritmic scale)
with respect to luminosities obtained from the $I_C$-band. 
This could help explaining the position of some sources above the locus of the
theoretical birthline.
Unknown binarity would also contribute to overestimate the 
luminosities and, consequently, to the shift of the sources above their real 
location in the HR~diagram. 

According to these evolutionary tracks, almost all of our sample sources
have masses between 0.8 and 2.0~$M_\odot$.
The mass of the interesting PMS~Ge object ABM\,22 appear to be somewhat above 
this value ($M\approx2.5~M_\odot$).
On the other hand, the bulk of the objects are younger than 
$\sim$ 3~Myr. ABM\,1 would, apparently, be much older but, as mentioned 
before, this star has very low quality 2MASS photometry 
and its location on the HR diagram is probably spurious. 
Thus, disregarding the latter, we find that 23 stars ($\sim65\%$) are younger 
than 1~Myr, 9 ( $\sim25\%$) have ages between 1 and 3~Myr and only 3 
($<10\%$) appear to be older than 3~Myr.

Based on the observed morphology and characteristics we could naturally 
distinguish three basic regions in M8: (i) a central part, roughly  
coincident with the young open cluster NGC~6530 itself, (ii) a brightest 
and probably more active part to the east of NGC~6530, known as the Hourglass 
nebula, and (iii) the southern edge formed by a rimmed nebula defined as 
``Southeastern Bright Rim'' 
and ``Extended Bright Rim'' by Lada et al. (1976). Tothill et al. (2002) 
resolved several continumm (at 850\,$\mu$m) and CO clumps in the last area, 
which contains a considerable number of the new PMS~objects. The dashed lines
in Figure~\ref{campo_m8} approximately delimitate the mentioned regions. 
We have analyzed the spatial distribution of the PMS~stars 
in relation to their ages. In order to quantify this distribution, we can
group the stars in two populations: a ``younger'' one, with age 
$t\lesssim1.5$~Myr
and an ``older'' one, with $t>1.5$~Myr. 
We find that the younger population 
is primarily located in the southern rim area (9 of 10 stars) and in 
the Hourglass region (10 of 11 stars). On the contrary, a mix of younger
and older PMS~stars (10 and 5, respectively) is observed in the central part. 

\section{Summary and Conclusions}

From the analysis of spectra obtained with the Boller \& Chivens spectrograph 
at the 6.5\,m Magellan I telescope (Las Campanas Observatory, Chile),
we identified 37 new PMS~objects among probable faint members 
of the young open cluster NGC~6530, at a distance of 1.25~kpc. 
Whereas 34 of them are classified as classical or weak-lined T~Tauri stars 
with spectral types K0-K7, 
we separated the 3 G-type objects in a different stellar class 
denominated ``PMS~Fe/Ge class'', as proposed by Mart\'{\i}n (1997).
The new PMS stars are more or less uniformly spread over 
the whole star forming region, with a marked 
concentration of YSOs toward the Hourglass Nebula. 

We studied the near-infrared properties of the newly discovered PMS stars 
using $JHK_s$ magnitudes either from 2MASS or from Arias et al. (2006). 
We found that their positions in the colour-colour diagram are compatible 
with those expected for T~Tauri and PMS~Fe/Ge stars affected by low to 
moderate reddening. Some of the objects 
also show substantial infrared excess emission.
In addition, we positionally matched the observed stars with the X-ray
sources discovered with the {\em Chandra} ACIS instrument and published by 
Damiani et al. (2004), finding that all but two of the sources included in the 
FOV of the {\em Chandra} observation are detected as X-ray sources. 

Based on our spectral classification along with the already mentioned 
near-infrared data 
we derived the effective temperatures and luminosities of these PMS stars 
and placed them in the HR~diagram. We used Palla \& Stahler (1999)
evolutionary tracks and isochrones to estimate their masses and ages.
We found that almost all of the new PMS stars 
have masses between 0.8 and 2.0~$M_\odot$.
One particular object, ABM\,22, seems to have a higher mass 
($\approx2.5~M_\odot$) and shows interesting evidence of a Herbig-Haro outflow 
in its spectrum. Furthermore the bulk of the objects ($\sim90\%$) appear to be 
younger than 3~million years. 
A not negligible number of sources located above the locus of the theoretical
birthline in the HR~diagram points in favor of the recently estimated
distance modulus of 10.5 or smaller (Prisinzano et al. 2005; 
Arias et al. 2006), in contrast to the previously accepted 11.25 
(e.g. van den Ancker et al. 1997; Sung et al. 2000),
since a larger distance would displace the stars completely out of the
expected region.

From the derived ages and spatial distribution of the individual PMS~stars 
we find that the youngest stars are preferentially located in the Hourglass
nebula, the region ionized by the O star Herschel\,36 (H\,36), 
as well as in the southern rim of the M8 nebula, where several molecular 
emission clumps have been previously detected (Tothill et al. 2002). 
On the other side, both younger and older PMS~stars are observed in the 
central area identifiable with the young open cluster NGC~6530. 
Although the sample size is relatively small, these results may be suggesting 
some kind of sequential process that would be in accordance with 
the model proposed by Lada et al. (1976) and Lightfoot et al. (1984), 
in which NGC~6530 formed first and triggered the star formation on the
periphery of the cavity created by its hot stars.
The model states that the stars of NGC~6530 may have triggered the formation 
of the O star 9\,Sgr, which subsequently caused the formation of the even
younger H\,36.
In their X-ray study of the region, Damiani et al. (2004) found evidence of an 
age gradient across the field from northwest to south, also suggesting a 
sequence of star formation events qualitatively similar to that found in 
earlier studies.
We conclude that most of the low-mass PMS~population identified in the present 
study seem to be coeval to the massive stars 9\,Sgr and H\,36, 
for which Sung et al. (2000) fitted an isochrone of age $\sim1.5$~million 
years, whereas the few oldest T~Tauri stars could have been formed in an
earlier stellar generation, coeval to the 4~Myr MS stars that populate the 
fainter end (Sung et al. 2000).

\section*{acknowledgements}

We thank the anonymous reviewer for many comments and suggestions that have 
improved this paper.
This publication makes use of data products from the Two Micron All Sky
Survey, which is a joint project of the University of Massachusetts and the
Infrared Processing and Analysis Center/California Institute of Technology,
funded by the National Aeronautics and Space Administration and the National
Science Foundation, and it is also based on observations made with ESO 
Telescopes at the La Silla Observatory under programme ID 2064.I-0559.
This research has made use of Aladin and the Simbad Database, operated at CDS,
Strasbourg, France.

Financial support from FONDECYT No. 1050052 and from PIP-CONICET No. 5697 are 
acknowledged by RHB and JIA, respectively.
JIA also thanks the Departamento de F\'{\i}sica of Universidad de La Serena 
for the use of their facilities and the warm hospitality. 
The authors gratefully thank the staff at LCO for kind hospitality during the 
observing run and Miguel Roth for his useful comments on the first version of 
this manuscript.

\section*{Appendix}
\label{append}

\begin{figure*}
\includegraphics[width=16cm, angle=180]{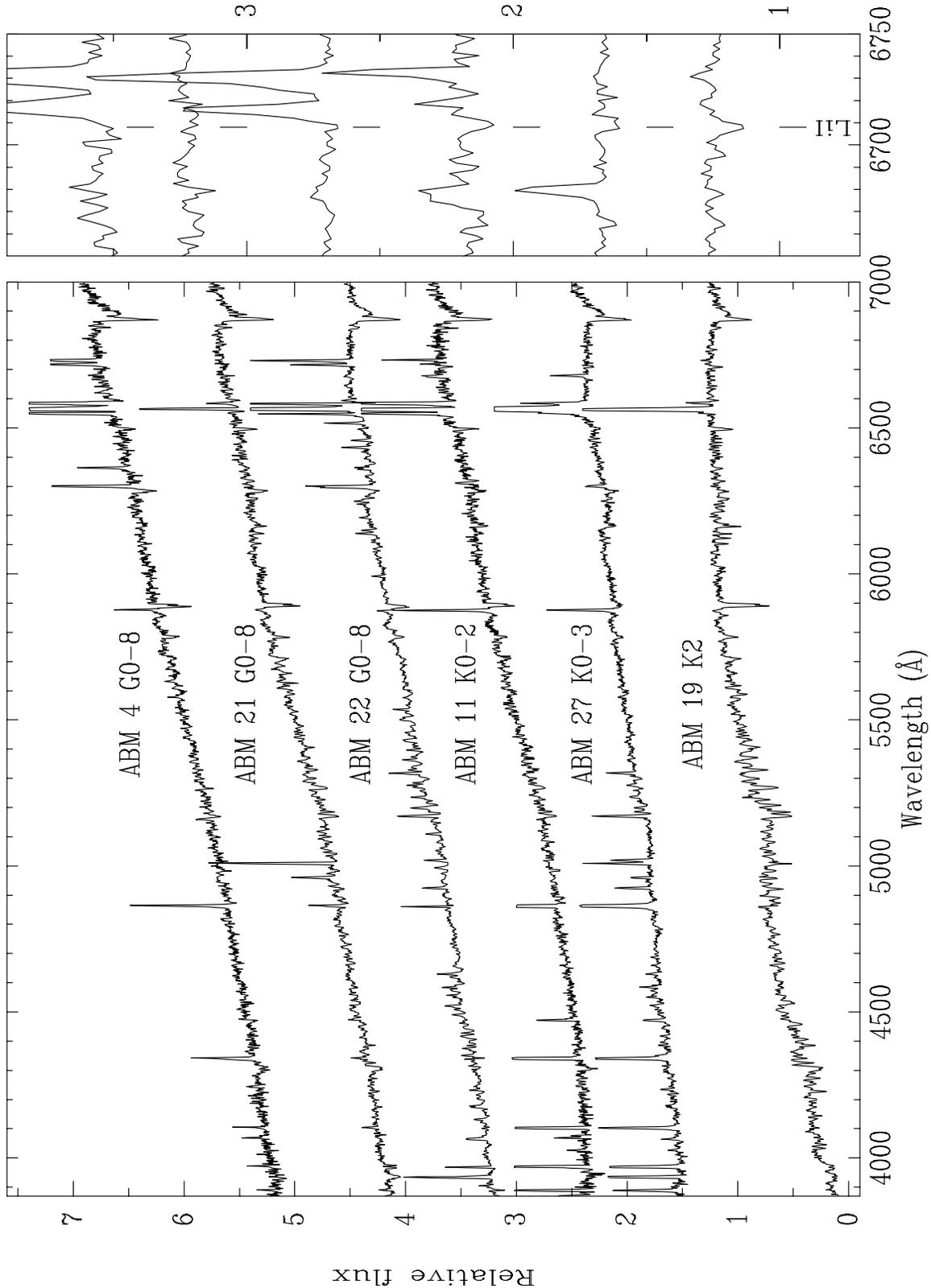}
\caption{Left-hand panel: spectra of the PMS~Fe/Ge and classical T~Tauri 
stars in M8 with spectral types G0-K2, covering the wavelenght range 
$3850-7000$\,\AA\,. The spectra have been normalized to the continuum at 
$\lambda=5500$\,\AA\,and vertically shifted for clarity. Some of the most 
prominent emission lines have been truncated with the same purpose. 
Right-hand panel: an enlargement of the same spectra showing the 
$6650-6750$\,\AA\,region. The ticks indicate the 
rest-wavelenght of the Li\,{\sc i}\,$\lambda$6708 line. }
\label{CTTS1}
\end{figure*}

\begin{figure*}
\includegraphics[width=16cm, angle=180]{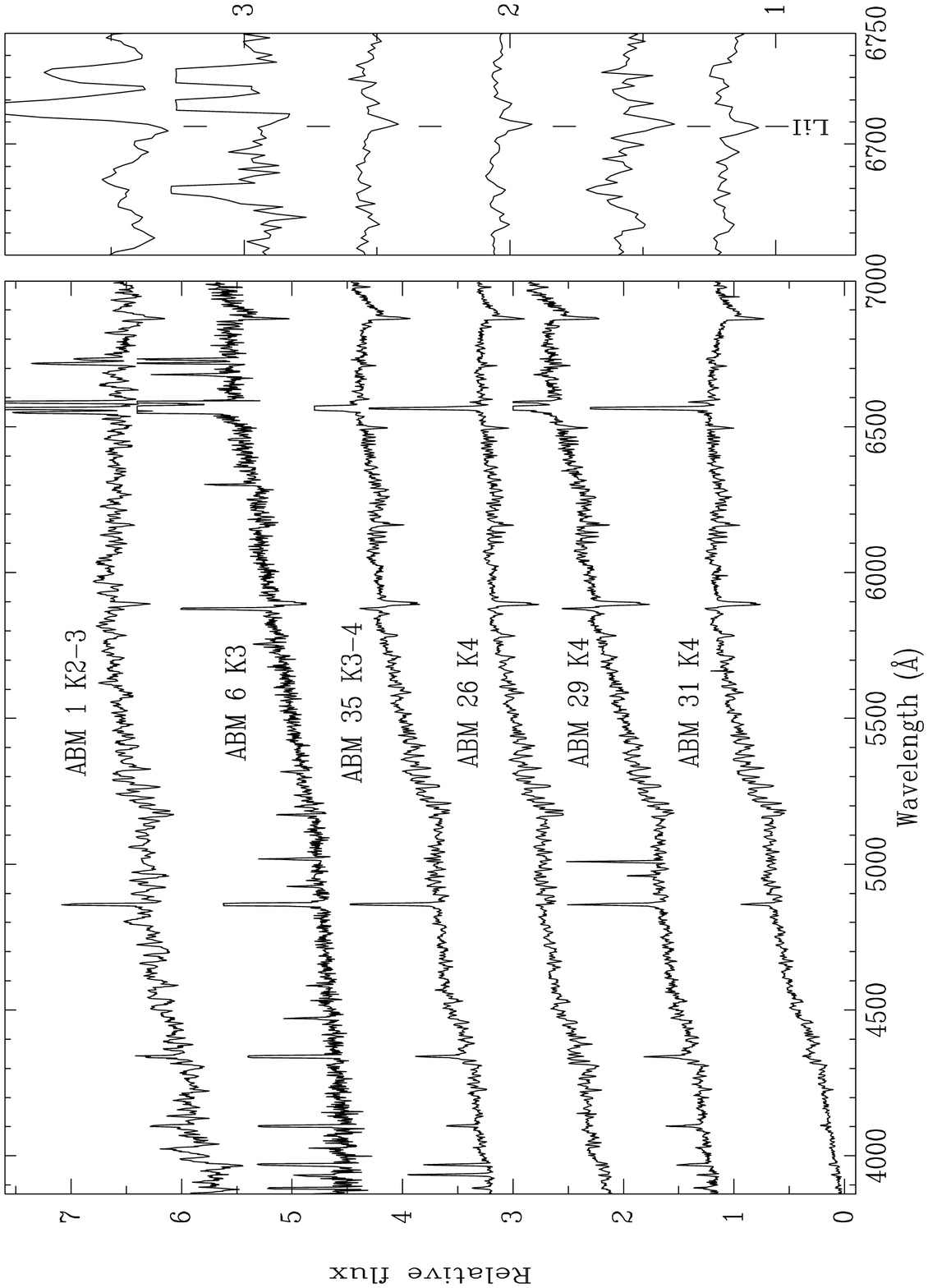}
\caption{Left-hand panel: spectra of classical T~Tauri stars in M8 with 
spectral types K2-K4, covering the wavelenght range $3850-7000$\,\AA\,. 
The spectra have been normalized to the continuum at 
$\lambda=5500$\,\AA\,and vertically shifted for clarity. Some of the most 
prominent emission lines have been truncated with the same purpose. 
Right-hand panel: an enlargement of the 
same spectra showing the $6650-6750$\,\AA\,region. The ticks 
indicate the rest-wavelenght of the Li\,{\sc i}\,$\lambda$6708 line.}
\label{CTTS2}
\end{figure*}

\begin{figure*}
\includegraphics[width=16cm, angle=180]{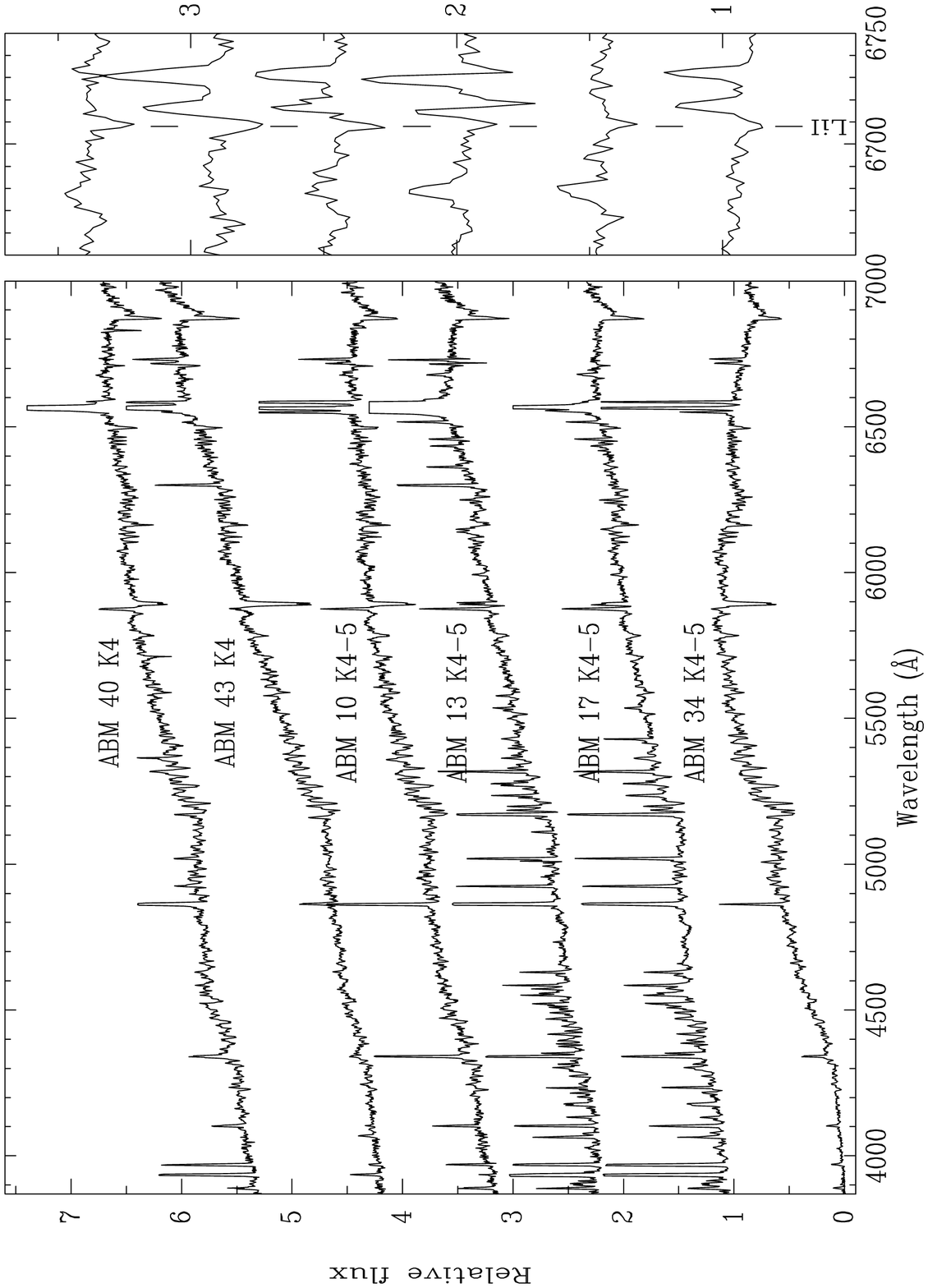}
\caption{Left-hand panel: spectra of classical T~Tauri stars in M8 with 
spectral types K4-K5, covering the wavelenght range $3850-7000$\,\AA\,. The 
spectra have been normalized to the continuum at $\lambda=5500$\,\AA\,and 
vertically shifted for clarity. Some of the most prominent emission lines 
have been truncated with the same purpose. Right-hand panel: an 
enlargement of the same spectra showing the $6650-6750$\,\AA\,region. The 
ticks indicate the rest-wavelenght of the 
Li\,{\sc i}\,$\lambda$6708 line.}
\label{CTTS3}
\end{figure*}

\begin{figure*}
\includegraphics[width=16cm, angle=180]{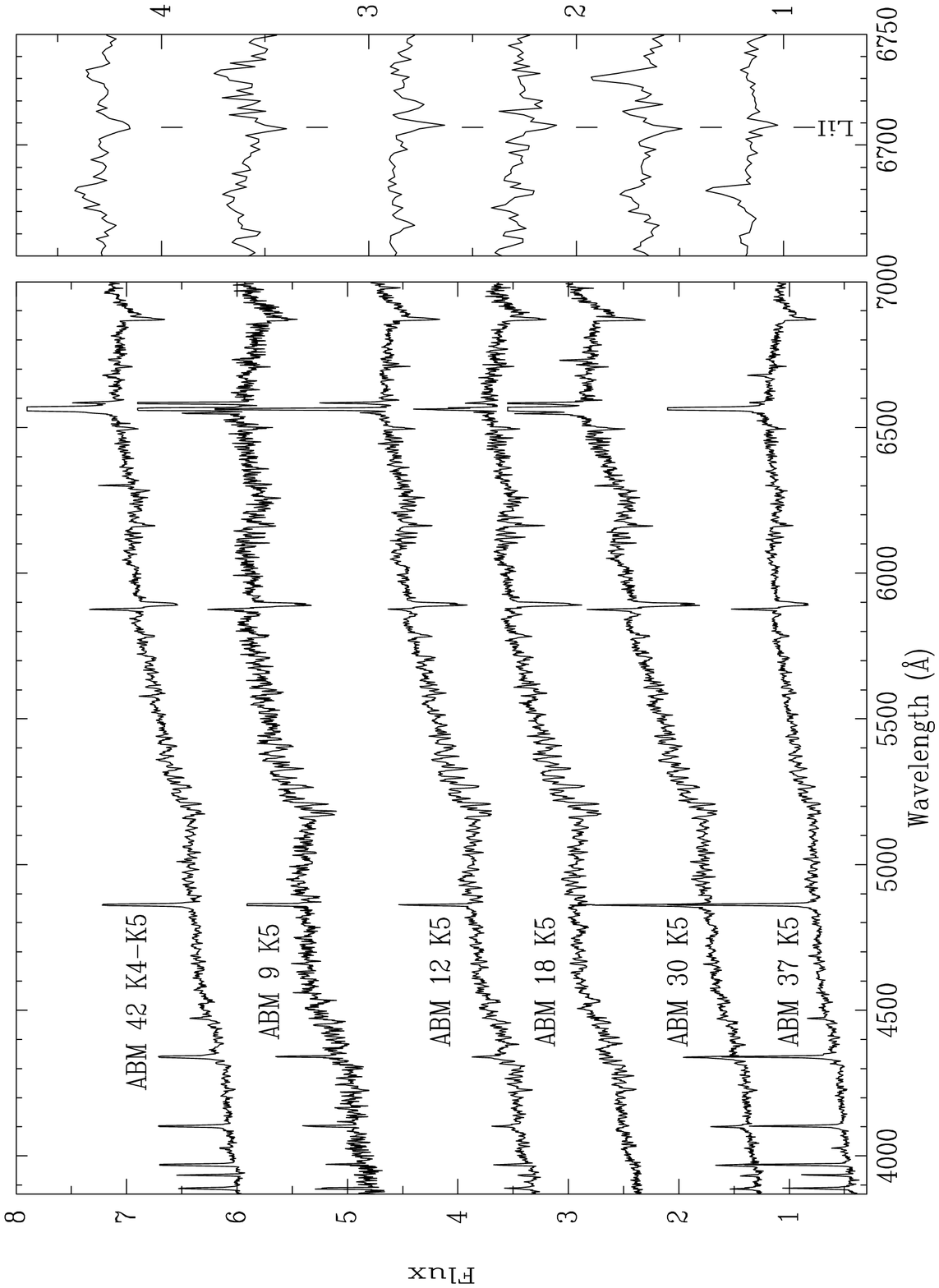}
\caption{Left-hand panel: spectra of classical T~Tauri stars in M8 with 
spectral type K5, covering the wavelenght range $3850-7000$\,\AA\,. The 
spectra have been normalized to the continuum at $\lambda=5500$\,\AA\,and 
vertically shifted for clarity. Some of the most prominent emission lines 
have been truncated with the same purpose. Right-hand panel: an 
enlargement of the same spectra showing the $6650-6750$\,\AA\,region. The 
ticks indicate the rest-wavelenght of the 
Li\,{\sc i}\,$\lambda$6708 line.}
\label{CTTS4}
\end{figure*}

\begin{figure*}
\includegraphics[width=16cm, angle=180]{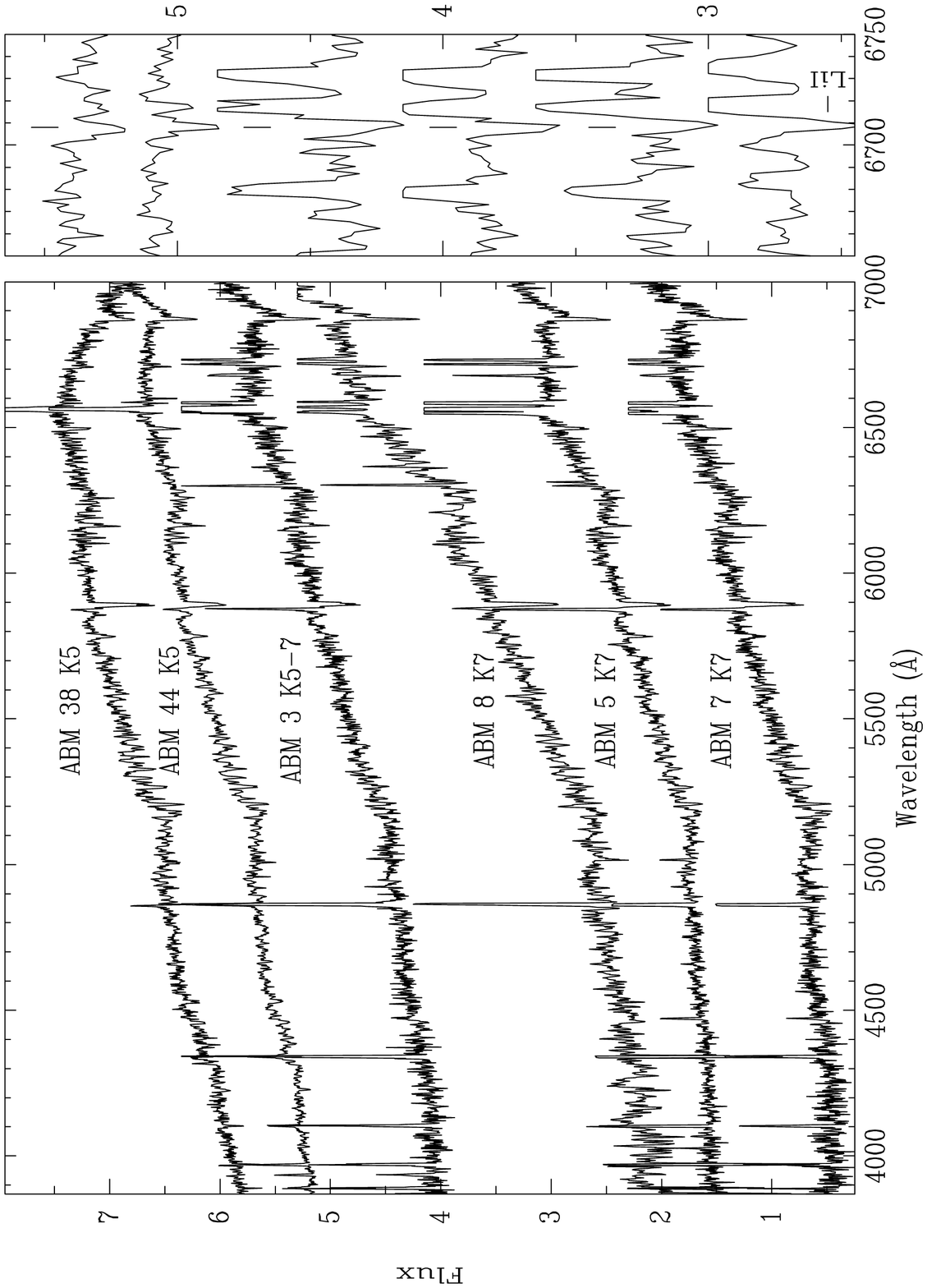}
\caption{Left-hand panel: spectra of classical T~Tauri stars in M8 with 
spectral types K5-K7, covering the wavelenght range $3850-7000$\,\AA\,. The 
spectra have been normalized to the continuum at $\lambda=5500$\,\AA\,and 
vertically shifted for clarity. Some of the most prominent emission lines 
have been truncated with the same purpose. Right-hand panel: an 
enlargement of the same spectra showing the $6650-6750$\,\AA\,region. 
The ticks indicate the rest-wavelenght of the Li\,{\sc i}\,$\lambda$6708
line.
}
\label{CTTS5}
\end{figure*}

\begin{figure*}
\includegraphics[width=16cm, angle=180]{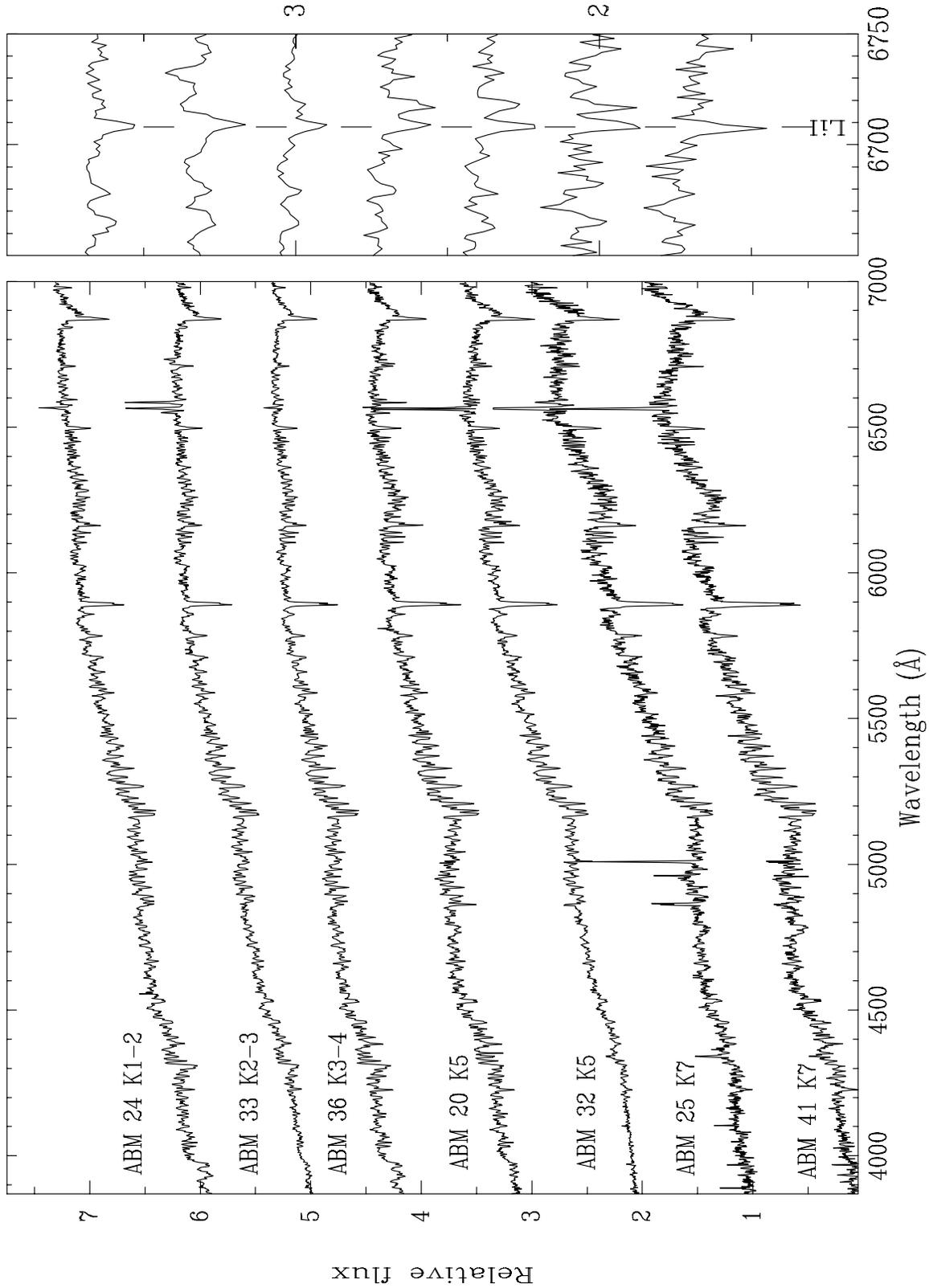}
\caption{Left-hand panel: spectra of weak-lined T~Tauri stars in M8, 
covering the wavelenght range $3850-7000$\,\AA\,. The spectra have been 
normalized to the continuum at $\lambda=5500$\,\AA\,and vertically shifted 
for clarity. Located in an extremely variable high-excitation region, the 
star ABM\,25 is about a hundred times fainter than the nebula in 
[O\,{\sc iii}], which impeded a perfectly accurate background subtraction 
in that part of the spectrum; thus, the lines observed at the 
[O\,{\sc iii}]\,$\lambda\lambda$4959, 5007 wavelenghts are not real emission 
lines but just residua of the processing.
Right-hand panel: an enlargement of the same spectra showing 
the $6650-6750$\,\AA\,region. The ticks indicate the rest-wavelenght of the 
Li\,{\sc i}\,$\lambda$6708 line.}
\label{WTTS}
\end{figure*}


\end{document}